\documentclass[fleqn,usenatbib]{mnras}

\usepackage{newtxtext,newtxmath}

\usepackage[T1]{fontenc}
\usepackage{adjustbox}
\DeclareRobustCommand{\VAN}[3]{#2}
\let\VANthebibliography\thebibliography
\def\thebibliography{\DeclareRobustCommand{\VAN}[3]{##3}\VANthebibliography}

\usepackage{graphicx}	
\usepackage{amsmath}	

\def\be{\begin{equation}}
\def\ee{\end{equation}}
\newcommand{\refeq}[1]{Eq.~(\ref{eq:#1})}          
\newcommand{\refeqs}[2]{Eqs.~(\ref{eq:#1})--(\ref{eq:#2})}  
\newcommand{\reffig}[1]{figure~\ref{fig:#1}} 
\newcommand{\reffigs}[2]{figures~\ref{fig:#1}--\ref{fig:#2}}
\newcommand{\refFig}[1]{Figure~\ref{fig:#1}}

\newcommand{\refsec}[1]{section~\ref{sec:#1}}

\newcommand{\reftab}[1]{table~\ref{table:#1}}

\def\Mpch{\, h^{-1} \, {\rm Mpc}}

\newcommand{\dbc}{\delta_{bc}}
\newcommand{\tbc}{\theta_{bc}}
\newcommand{\bdbc}{b_{\delta_{bc}}}

\def\prd{Phys.~Rev.~D}

\def\mnras{MNRAS}
\def\jcap{JCAP}
\def\apj{Astrophys.~J.}
\def\apjl{Astrophys.~J.~Lett.}

\usepackage[utf8]{inputenc}

\title{Cosmic voids and BAO with relative baryon-CDM perturbations}
\author[H.~Khoraminezhad et al.]{\parbox{\textwidth}{Hasti Khoraminezhad,$^{1,2,3}$\thanks{\texttt{Email:hkhorami@sissa.it}} 
Pauline Vielzeuf,$^{1,2}$ 
Titouan Lazeyras,$^{1,2,3}$ 
Carlo Baccigalupi $^{1,2,3,4}$ 
and Matteo Viel $^{1,2,3,4}$}
\vspace{0.3cm} \\
$^{1}$SISSA, Via Bonomea 265, 34136 Trieste, Italy\\
$^{2}$IFPU, Institute for Fundamental Physics of the Universe, via Beirut 2, 34151, Trieste, Italy\\
$^{3}$INFN, Sezione di Trieste, Via Bonomea 265, 34136 Trieste, Italy\\
$^{4}$INAF, Osservatorio Astronomico di Trieste, Via Tiepolo 11, 34143 Trieste, Italy 
}

\date{Accepted XXX. Received YYY; in original form ZZZ}

\pubyear{2021}

\begin{document}
\label{firstpage}
\pagerange{\pageref{firstpage}--\pageref{lastpage}}
\maketitle

\begin{abstract}
We study the statistics of various large-scale structure tracers in gravity-only cosmological simulations including baryons and cold dark matter (CDM) initialized with two different transfer functions, and simulated as two distinct fluids. This allows us to study the impact of baryon-CDM relative perturbations on these statistics. In particular, we focus on the statistics of cosmic voids, as well as on the matter and halo real-space 2-point correlation function and baryon acoustic oscillations (BAO) peak. We find that the void size function is affected at the 1-2\% level at maximum, and that the impact is more important at higher redshift, while the void density profile and void bias are roughly unaffected. We do not detect a sizeable impact of relative baryon-CDM perturbations on the real-space correlation functions of matter and halos or the BAO peak, which is in line with results from previous works. Our results imply that it would be hard to use voids or real-space correlation functions to constrain baryon-CDM relative perturbations, but also that we might not have to include them in models for the analysis of future cosmological surveys data.
\end{abstract}



\section{Introduction}
\label{sec:intro}

The different evolution of baryons and cold dark matter (CDM) due to photon pressure before recombination causes relative perturbations between the two fluids in the early Universe. These perturbations can be both in the density and peculiar velocity of the two fields but, importantly, they keep the total matter perturbations unchanged, and are thus referred to as \textit{relative baryon-CDM density perturbations} and \textit{relative baryon-CDM velocity perturbations} (\cite{Dalal_2010,Tseliakhovich_2010,Yoo_2011,Barkana:2010zq,Yoo:2013qla,Slepian:2014dda,Blazek:2015ula,Slepian:2016nfb,Schmidt:2016,Beutler:2016zat,Khoraminezhad:2020zqe}). After recombination, these primordial relative perturbations are slowly erased by gravitational evolution with baryons falling in CDM potential wells. In standard studies of Large-Scale Structure (LSS), this process is assumed to be complete before redshift zero, and baryons and CDM are treated as one single comoving matter fluid. However, this assumption is not exactly correct, and there were several recent efforts to describe the evolution of baryons and CDM as two distinct fluids across cosmic history (see in particular \cite{Tseliakhovich_2010, Barkana:2010zq, Schmidt:2016, Beutler:2016zat, Chen:2019cfu, Rampf:2020ety}). Notice that similar perturbations can also be generated in some inflationary scenarios, and are then referred to as Compensated Isocurvature Perturbations (CIPs) (\cite{Polarski:1994rz,Linde:1996gt,Liddle:1999pr,Langlois:2000ar,Notari:2002yc,Lyth:2002my,Ferrer:2004nv,Li:2008jn,Grin:2011,Valiviita:2012ub,Huston:2013kgl,Christopherson:2014eoa,He:2015msa,Heinrich:2019sxl,Barreira:2020lva}). However, in this work, we do not treat these CIPs, and we focus only on relative baryon-CDM perturbations induced by photon pressure prior to recombination. 

\textit{2-fluid simulations} in which baryons and CDM are initialized with two different transfer functions and are considered as two distinct fluids coupled gravitationally are starting to play an important role in this line of study (see \cite{Yoshida:2003,OLeary:2012,Angulo:2013qp,Bird:2020,Hahn:2020lvr,Michaux:2020yis,Khoraminezhad:2020zqe}). Crucially, these are gravity-only simulations (i.e. they do not include any late-time hydrodynamics), and early-Universe baryonic effects only enter through the use of different transfer functions to initialize baryons and CDM.   

Relative velocity perturbations were identified for the first time by \cite{Tseliakhovich_2010}, while relative density perturbations were first pointed out in \cite{Barkana:2010zq}. In both cases, they are expected to affect structure formation (\cite{Ahn:2016bcr}), as well as the clustering of LSS tracers (\cite{Schmidt:2016,Beutler:2016zat,Barreira:2019qdl,Khoraminezhad:2020zqe}). This is because the coupling of baryons to photons before recombination prevents baryons from evolving gravitationally together with CDM, and consequently acts against structure formation and clustering, an effect that might need to be taken into account in studies of LSS. The formalism to include baryon-CDM relative perturbations in the statistics of LSS tracers was first discussed in \cite{Schmidt:2016} using the bias formalism (see \cite{Desjacques:2016bnm} for a complete review on this formalism). The main point is the need to add new terms proportional to these relative perturbations to the bias expansion, which links the density of tracers such as halos or galaxies $\delta_h$ to various underlying perturbations. At linear order, these terms consist of the relative density perturbation $\dbc$ (with $\dbc=\delta_b-\delta_c$) and relative velocity divergence perturbation $\tbc$(with $\tbc=\theta_b-\theta_c$) multiplied by their respective bias parameters $b_{\delta_{bc}}$ and $b_{\theta_{bc}}$, and the overdensity of halos can be written as (note that here $\mathbf{x}$, indicates the Eulerian position)  
\be
\delta_h (\mathit{\mathbf{x}},z) = b_1(z) \delta_m(\mathbf{x},z) + b_{\delta_{bc}}(z) \delta_{bc}(\mathbf{x}) + b_{\theta_{bc}}(z) \theta_{bc}(\mathbf{x},z),
\label{eq:bias-expansion}
\ee
where $b_1$ is the standard linear bias. The parameters $\bdbc$ and $b_{\tbc}$ were studied in previous works (see for example \cite{Barkana:2010zq,Schmidt:2016,Beutler:2016zat,Barreira:2019qdl,Chen:2019cfu,Hotinli:2019wdp,Khoraminezhad:2020zqe}). Specifically, \cite{Barreira:2019qdl} used the separate universe simulations technique to do the first measurement of $b_{\delta_{bc}}$ (corresponding to CIPs generated during Inflation), while \cite{Khoraminezhad:2020zqe} measured $b_{\delta_{bc}}$ using gravity-only 2-fluid simulations (corresponding to relative perturbations generated by photon pressure), and showed the two parameters to be equal. This work is a follow-up of \cite{Khoraminezhad:2020zqe}, and we will investigate the effects that such perturbations could induce on specific structures and cosmological probes. It is worth mentioning that one of the first usage of the separate universe technique for isocurvature perturbations appeared in \cite{Jamieson:2018biz}. This was done for the case of dark energy/CDM relative perturbations but is nevertheless somewhat related to the perturbations we consider here, and pioneered the use of separate universe simulations for isocurvature perturbations.

The first structures we consider are cosmic voids. Cosmic voids are defined as large underdense regions of the cosmic web, they are the largest structures in the Universe and make up most of its volume (\cite{10.1093/mnras/stu768,10.1093/mnras/stv879}). Historically, their existence was one of the earliest predictions of the concordance cosmological model (\cite{Hausman_1983}), and their observational detection goes back to roughly 40 years ago (\cite{Gregory_1978,krishner_1981}). Voids are in particular extremely underdense near their centers, and their spherically averaged density profile shows a characteristic shape (\cite{Colberg:2004nd,PhysRevLett.112.251302,10.1093/mnras/stu307,10.1093/mnras/stt1069,Nadathur2016b}). Recently, cosmic voids are becoming a promising cosmological probes: firstly they could represent a population of statistically ideal spheres with a homogeneous distribution at different redshifts which size evolution could be used to probe the expansion of the Universe using  Alcock \& Paczynski tests (\cite{Alcock_1979,Lavaux_2012,Sutter_2012,10.1093/mnras/stu1392,Hamaus:2015yza,Hamaus:2016wka,Mao:2016onb,hamaus2021euclid}). Moreover, due to their low density, voids are naturally sensitive to dark energy and thus the interest to use them as probe of alternative Dark Energy models and modified gravity scenarios is increasing (\cite{PhysRevD.92.083531,PhysRevD.80.103515,PhysRevD.83.023521,Bos:2012wq,10.1093/mnras/stt2298,Lavaux_2010,10.1093/mnras/stt219,10.1111/j.1365-2966.2010.17867.x,Cai:2014fma,Barreira:2015vra,10.1093/mnras/stv1209,10.1093/mnras/stv2503,Baldi:2016oce}), as well as the possibility of using them to put constraints on neutrinos masses (\cite{Massara:2015,10.1093/mnras/stz1944,Contarini_2021}). Their imprint on the observed Cosmic Microwave Background (CMB) is also becoming an encouraging new cosmological probe, either through their Integrated Sachs-Wolfe (ISW) imprint (\cite{Baccigalupi_1997,Baccigalupi_1999,granett2008,Cai2014,Hotchkiss2015,Planck2015,granettt2015,Nadathur2016,kovacs2017,kovacs2019,hang2021}), or their lensing imprint (\cite{cai2017,Raghunathan2020,vielzeuf2021}). Furthermore, the observed cold spot of the CMB could be explained as the imprint of the ISW sourced by very large voids along the line of sight (\cite{Rees_1968,finelli_2014,10.1093/mnras/stt2241,PhysRevD.90.103510}). Moreover, some works such as \cite{Jamieson:2019dmp} studied the properties of the voids via the separate universe technique. Finally, some studies tried to link high redshift intergalactic voids in the transmitted Lyman-$\alpha$ flux to the gas density (\cite{Viel_2008}). Because they are almost empty regions, their evolution during cosmic history is at most weakly nonlinear and their properties could possibly be impacted by the primordial density fields from which they formed. This fact motivates us to investigate the effects of baryon-CDM relative perturbations on these objects and their statistics.

Second, we will consider the real-space correlation function of various fields in our simulations. We will in particular focus on the Baryon Acoustic Oscillation (BAO) feature. Measuring the BAO feature in the distribution of galaxies is one of the most powerful tools for precision cosmology. For instance, the latest cosmological implications from final measurements of clustering using galaxies, quasars and Ly$\alpha$ forests from the Sloan Digital Sky Survey (SDSS) reported the following cosmological constraints: $H_0 =68.20 \pm 0.81\,\, \rm{km}\, \rm{s^{-1}} \rm{Mpc^{-1}}$ and $\sigma_8=0.8140 \pm 0.0093$ allowing for a free curvature and a time evolving equation of state for the dark energy (\cite{eBOSS:2020yzd}). Furthermore, combining the full-shape and BAO analyses of galaxy power spectra of the final Baryon Oscillation Spectroscopic Survey (BOSS) data release, \cite{Philcox:2020vvt}, recently obtained a $1.6\%$ precision measurement of $H_0$. Recent works suggest that relative baryon-CDM perturbations $\delta_{bc}$ and $\theta_{bc}$ could provoke possible systematics in the estimation of the BAO peak position (\cite{Dalal_2010,Yoo:2013qla,Barkana:2010zq,Schmidt:2016,Beutler:2016zat,Barreira:2019qdl}), and thus could potentially bias the cosmological constraints as a systematic shift in $D_{A}(z)$, $H(z)$, and $f\sigma_8$ measurements. 

The goal of this paper is to assess the impact of relative baryon-CDM perturbations on one side cosmic voids, and on the other side on the real-space correlation functions of various fluids, in particular the position of the BAO peak. We do this using the aforementioned 2-fluid simulations, and compare the results with those obtained in a standard gravity-only 1-fluid simulations. We emphasis that we work in configuration space, in contrast with our first paper where we worked in Fourier space (\cite{Khoraminezhad:2020zqe}). We first give a detailed description of our numerical arrangement in \refsec{num}, including details of our simulation setup and the halo finding procedure (\refsec{sims}), as well as the void finding algorithm (\refsec{voidfind}). We then investigate the impact of baryon-CDM perturbations on the void size function (VSF) using different tracers of the underlying matter field to identify cosmic voids (namely particles and halos) in \refsec{voids}. In \refsec{void2pcf}, we measure the void-void and halo-void correlation functions (\refsec{voidfull}), the void density profile (\refsec{voidprofile}), and the void bias (\refsec{voidbias}) in presence of baryon-CDM perturbations. We further explore the effect of such perturbations on the real-space matter and halo 2-point correlation functions (2PCF) in \refsec{bao}, and in particular, we compare the position of the BAO peak in the 2PCF of total matter, halos, CDM, baryons and the relative density $\delta_{bc}$ in \refsec{peak}. Finally, we draw our conclusions in \refsec{conclude}. 

\section{Numerical setup}
\label{sec:num}

\subsection{Simulations and halo finding}
\label{sec:sims}

Our $N$-body simulation suite is based on the one presented in \cite{Khoraminezhad:2020zqe}, and consists of
\begin{enumerate}
\item a set of collisionless gravity-only simulations in which baryons and CDM are evolved as two distinct fluids initialized from two distinct primordial power spectra as predicted by early universe physics. We refer to this set of simulations as ``2-fluid''.
\item a set of a standard gravity-only simulation in which the baryons and CDM are considered as perfectly comoving and are hence simulated as one total matter field. We refer to this set as ``1-fluid''. 
\end{enumerate}
Our cosmology is consistent with Planck 2018 (\cite{Aghanim:2018eyx}) $\Lambda$CDM, namely: $\Omega_{m}=0.3111$, $\Omega_{b}=0.0490$, $\Omega_{c}=0.2621$, $\Omega_{\Lambda}=0.6889$, $n_{s}=0.9665$, $\sigma_{8}=0.8261$ and  $h=0.6766$. In this work, we enlarge our previous simulation box size to $L_{\rm box}=500 \, \Mpch$ on each side to be large enough for void finding. We perform 8 realizations of each types of simulations (1-fluid/2-fluid) with $512^3$ particles of each species. Importantly, each realisation was initialized with a different random seed but the seeds used for total matter in 1-fluid simulations are the same as the ones used for CDM in 2-fluid ones in order to minimize cosmic variance. The details of the simulations are given in \reftab{simu}. 

To generate the initial conditions of the density and velocity perturbations we used the publicly available initial condition code ``MUSIC'' (\cite{Hahn_2011}) at redshift $z_i=49$. For the 1-fluid case, we compute the matter power spectrum at $z=0$ using the publicly available Boltzmann code CAMB (\cite{Lewis_2000}) and back-scale it to the initial redshift, while for the 2-fluid scenario we use the two different transfer functions for baryons and CDM directly at $z_i=49$. We use the first order Lagrangian perturbation theory, Zel'dovich approximation, (\cite{Zeldovich-1970})  to estimate the velocity as well as the density fields. In order to reduce the effect of cosmic variance, we use the fixed-mode amplitude technique implemented in the MUSIC code (\cite{Angulo:2016hjd}). Importantly, we keep the total matter power spectrum the same for the 1-fluid and 2-fluid scenarios, and we use the same random seeds to initialize 1-fluid simulations and CDM particles in the 2-fluid case.

We perform our simulations using the cosmological N-body code GADGET-II (\cite{Springel:2005mi}). In the case of 2-fluid simulations, as was discussed in \cite{Angulo:2013qp,Khoraminezhad:2020zqe}, since we have two different fluids (baryons as the light fluid and CDM as the heavy one), and a too high force resolution for a given mass resolution would lead to a spurious coupling between baryons and CDM, we use adaptive gravitational softening (AGS) (\cite{Iannuzzi2011AdaptiveGS}) for baryons only, which allows the softening length to vary in space and time according to the local density, and alleviates the spurious coupling arising between CDM and baryon particles, as was discussed in \cite{Angulo:2013qp,Khoraminezhad:2020zqe}. In more details, in the 2-fluid simulation suite, the force affecting baryonic particles is softened adaptively using an SPH kernel with a size set by the $14^{\rm th}$ closest neighbours. Moreover we set the floor minimum softening length $\epsilon = 25 h^{-1} \rm{kpc}$, which corresponds to $1/40$-th of the mean interparticle separation of the baryons. We note that the CDM softening length is kept constant through space and time to $\epsilon = 25 h^{-1} \rm{kpc}$, which corresponds to the $1/40$-th of the mean CDM interparticle separation as well. These settings are tested and validated in details in section 3.3 and Appendix B of \cite{Khoraminezhad:2020zqe}. Finally, we insist again that since we are interested in computing the effect of \textit{early} baryon-CDM perturbations on LSS, we neglect the late-time impact of baryonic processes and do not include hydrodynamical forces in the simulations. We refer the reader to \cite{Khoraminezhad:2020zqe} for all the details and validating tests of our numerical setup. 

\begin{table}
\centering
\resizebox{8.5cm}{!}{
\begin{tabular}{|c|c|c|c|c|c|c|} 
 \hline
 Name  & $L_{\rm box}$ & $N_b$ & $N_c$ & $m_b$ & $m_c$ & $N_{\rm real}$  \\ 
\tiny{--} & \tiny{$[{\rm Mpc}/h]$} & \tiny{--} & \tiny{--} & \tiny{$[10^{10} M_\odot/h]$} & \tiny{$[10^{10} M_\odot/h]$} & \tiny{--}\\
 \hline \hline
 1-fluid & $500$ & 0 & $512^3$ & -- & $1.0051$ & 8  \\ 
 \hline
 2-fluid & $500 $ & $512^3$ & $512^3$ & $0.1583$ & $0.8468$ & 8  \\ 
 \hline
\end{tabular}
}
\caption{Principal parameters of our numerical setup. $L_{\rm box}$ denotes the length of the side of the box, $N_b$ and $N_c$ are the number of baryon and CDM particles respectively, $m_b$ and $m_c$ denote their corresponding mass in units of $10^{10} M_\odot/h$, and $N_{\rm real}$ is the number of realizations.}
\label{table:simu}
\end{table}

We use the spherical overdensity (SO) algorithm Amiga Halo Finder (AHF) (\cite{Gill:2004,Knollmann:2009}) to identify halos. The definition of the virial radius is the one of a sphere  in which the average density is given by $\bar{\rho}_{vir}(z)= \Delta_{m}(z) \, \rho_{m}(z)$ where $\rho_m$ is the background total matter density. We chose the overdensity threshold as $\Delta_m=200$, and set the minimum number of particles per halo to $20$. For this work, we only used main halos and discarded subhalos from the catalogues. We identify halos at redshift $z=0, \, z=0.5, \, z=1, \, z=1.5, \, z=2,$ and $z=3$. In the case of 2-fluid simulations, we use both CDM and baryon particles to identify halos. We compared the halo mass function in the 1-fluid and 2-fluid simulations and found good agreement (see figure 3 of \cite{Khoraminezhad:2020zqe}). 

\subsection{Void Finder}
\label{sec:voidfind}

We use the publicly available REVOLVER (REal-space VOid Locations from surVEy Reconstruction)\footnote{https://github.com/seshnadathur/Revolver}  void finder to build our void catalogues with the ZOBOV (ZOnes Bordering On Voidness) algorithm (\cite{ZOBOV}), which is a 3D void finder and has been widely used both in simulations and observed catalogues (\cite{DES:2021gua,Nadathur:2020vld,Contarini_2021,Nadathur:2020vld}). The ZOBOV algorithm performs a Voronoi tessellation of a set of points, identifies depressions in the density distribution of these points, and merges them into group of Voronoi cells using a watershed transform \citep{Platen:2007qk} without pre-determined assumptions about voids shape, size or mean underdensity, which is the most appealing aspect of the watershed method. Here we briefly outline the basic steps of the void-finding technique in ZOBOV and we refer the interested readers to the main ZOBOV paper (\cite{ZOBOV}) for a detailed description. One can describe the ZOBOV mechanism with the four following main steps:
\begin{enumerate}
    \item \textit{Voronoi Tessellation Field Estimator (\cite{Schapp-2007}}): the algorithm divides the space into cells around each tracer $i$ (halos or particles in this work) in which the region inside the cell is closer to particle $i$ than to any other one. It then estimates the density of each Voronoi region using the volume of each cell $1/V(i)$. 
    \item \textit{Definition of the minimum density}: after estimating the density in each cell in the first step, the algorithm finds the minimum density cells, defined as Voronoi cells with a density lower than all their neighboring ones. 
    \item \textit{Formation of basins}: the algorithm then joins adjacent higher-density cells to the minimum-density cell until no neighbor cell with a higher density can be found. It means that the void finder links all the particles to their minimum density neighbour. This procedure defines basins as the zones of these cells. At this point, these basins themselves could be considered as voids because they are depression regions in the density field, but one single basin may also arise from spurious Poisson fluctuations due to the discreteness of the particles.
    \item \textit{Watershed transform}: the last step is when these basins are joined together using a watershed algorithm (\cite{Platen:2007qk}). For each basin $b$, the ``water'' level is set to the minimum density of $b$. It is then slowly elevated so that it can flow to the neighbor basins, joining all of them to basin $b$. The process stops when the ``water'' flows into a basin with a lower minimum, which defines the final void distribution.
\end{enumerate}

Void centers are then defined as the center of the largest sphere completely empty of tracer that can be inscribed within the void. Indeed, this is the best predictor of the location of the minimum of the matter density field (\cite{Nadathur:2015qua}). The effective radius of the void, $R_v$, is computed using the total volume of the underdense region and assuming sphericity 
\begin{align}
    V_{\rm void} \equiv \sum_{i=1}^{N} V_i^t = \dfrac{4}{3} \pi R_{v}^3,
    \label{sphere}
\end{align}
where $V_i^t$ is the volume of the Voronoi cell of the $i$th tracer, and $N$ represents the number of points that are included in the void.

We run the ZOBOV algorithm for all realizations of our 1-fluid and 2-fluid simulations presented in \refsec{sims} at redshift $z=0, \, z=0.5, \, z=1, \, z=1.5, \, z=2,$ and $z=3$ for two tracers:
\begin{itemize}
    \item Halos 
    \item Dark matter particles.
\end{itemize}
In order to better handle the computational cost of running the void finder in the particle field, we have made a down-sampling routine to randomly select CDM particles of the simulation snapshots down to a constant average density of $6.71 \times 10^6$ particles per cubic box-size ($500 \Mpch$), which corresponds to $5\%$ of the particles at each redshift, and insures us to be conservative with the density. We have verified that the different void statistics we study here were not affected when using a different random sample. We note that in the case of the 2-fluid simulation scenario, even if we have both types of particles (baryons \& CDM) in the simulation, we only used the down-sampled positions of CDM particles. We should in principle select voids in the total matter density field, including baryons, however, the ZOBOV algorithm can not discriminate between different populations of particles with different masses. Therefore we must identify the voids in one of the two density fields only. Since CDM particles are much more massive than baryons, they are more representative of the underlying total mater field, and are the stronger contributor to the evolution of cosmic structures. We emphasize that we do not expect the inclusion of baryons or not in the void finding procedure to strongly affect our results.   

We note that the total number of voids identified in the particle-field is significantly greater (from $\sim 20$ times for $z=0$ to $ \sim 200$ times for $z=3$) than the number of voids in the halo field due to the difference in the mean tracer densities (\cite{10.1093/mnras/stz1944}). Moreover, for both types of simulations when one uses halos as tracer of the matter field, the total number of voids gradually decreases with increasing redshift (for instance for the $1^{\rm st}$ realization of our 1-fluid simulation we found $2085$, $1950$, $1621$, $1225$, $860$ and $289$ voids at $z=0$, $z=0.5$, $z=1$, $z=1.5$, $z=2$ and $z=3$ respectively) which is due to the fact that the number of halos formed at higher redshift is smaller than the ones at lower redshift which decreases the tracer density at higher redshift, and consequently the number of voids. On the other hand, in the case where CDM particles are used as tracer, the total number of voids increases as the redshift increases since we kept the tracer density constant at all redshift in this case (for example for the same $1^{\rm st}$ realization of the 1-fluid simulation in the particle field we found $32544$, $42208$, $52188$, $61642$, $70076$ and $83430$ voids at $z=0$, $z=0.5$, $z=1$, $z=1.5$, $z=2$ and $z=3$ respectively). In order to understand these features in the statistics of the voids in a better way, we will look at the distribution in size of cosmic voids in the next section.

\section{Void Size Function}
\label{sec:voids}

The Void Size Function (VSF), or void abundance (\cite{Sheth2004,Furlanetto2006}) is the number of voids in a given radius bin at a given redshift. The VSF is a relatively recent tool that nowadays is becoming promising to probe dark energy (\cite{pisani2015,Verza2019}) as well as constraining neutrino masses (\cite{Massara:2015,10.1093/mnras/stz1944,Contarini_2021}). In addition to that, some recent works have also explored the differences between VSF in the concordance model of cosmology $\Lambda$CDM and modified gravity theories (see \cite{Cai:2014fma}), Galileon or non-local gravity (see \cite{Barreira:2015vra}), or the possibility of couplings between CDM and dark energy (see \cite{10.1093/mnras/stv2503}). Here we will present the comparison between the VSF in 1-fluid and 2-fluid simulations to assess the impact of baryon-CDM relative perturbations on these statistics. Each time we focus on voids identified both in the CDM density field (down-sampled) and in the distribution of collapsed halos to highlight how the use of different tracers with different bias might result in a different relative behaviour in the VSF. Notice that the impact of these perturbations has been studied in \cite{Khoraminezhad:2020zqe} for key observables of overdense regions of the density field (halo mass function and power spectrum, and the contribution of the baryon-CDM perturbation bias term to the halo power spectrum was found to be at maximum $0.3 \%$ at $k=0.1 \, \rm{h\,Mpc^{-1}}$, at $z=0$). However, they remain unexplored for underdense regions observables. 

\begin{figure*}
\centering
\includegraphics[width=8cm]{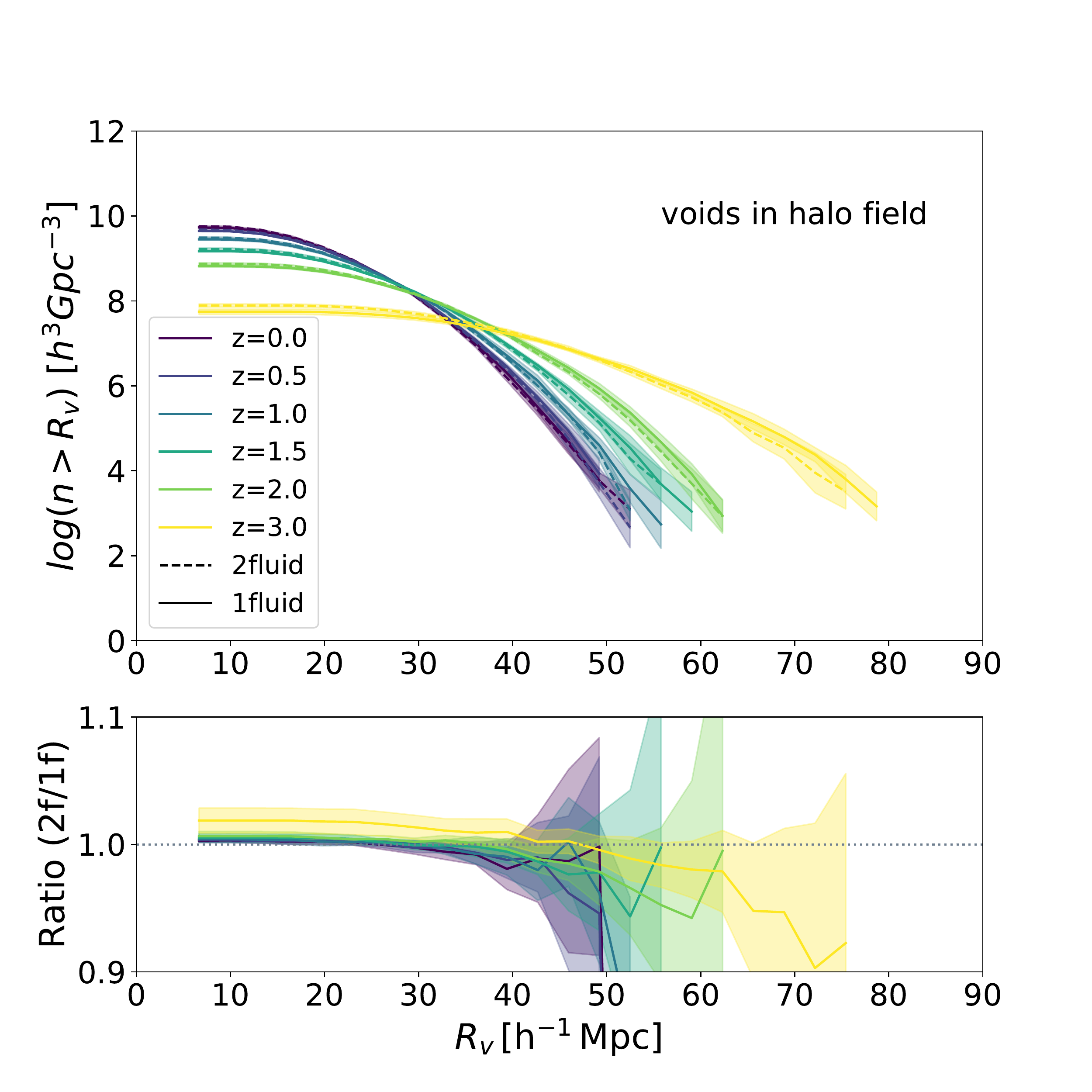}\hfill
\includegraphics[width=8cm]{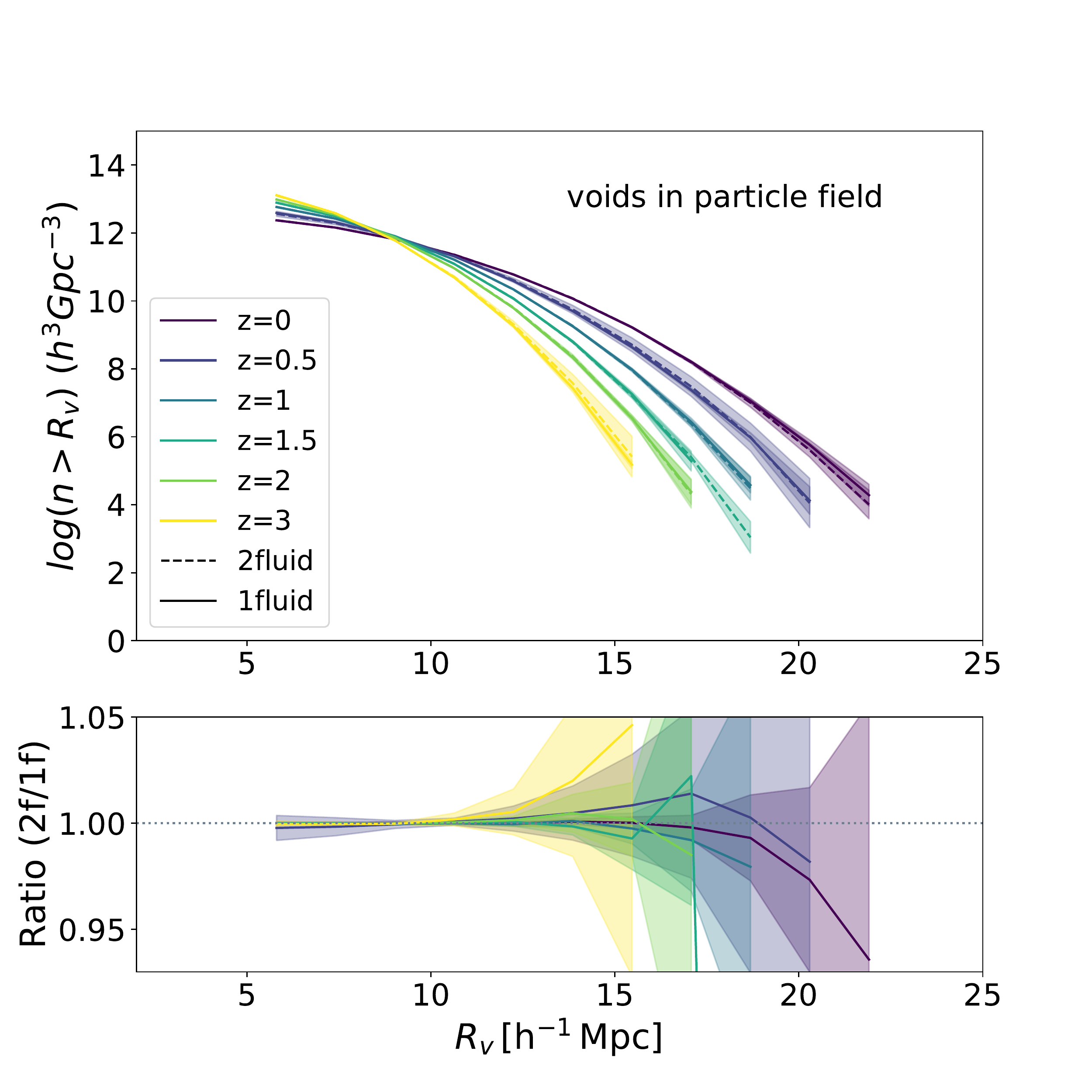}\hfill
\caption{Cumulative void size function (number density of voids with radii above $R_v$) in the 2-fluid simulations in dashed and 1-fluid simulations in solid lines in the \textbf{halo field} (left panel) and in the \textbf{particle field} (right panel) for different redshift illustrated by the color bar. The lower panels show the ratio of the VSF as ``2-fluid/1-fluid'' to see the difference better. The gray dotted line in the lower panels stand for the situation in which the VSF is equal in both types of simulations. The shaded area in each case depicts the $1\sigma$ error on the mean obtained from the $8$ different realizations.}
\label{fig:VSF_halo_particle}
\end{figure*}

\subsection{VSF in the halo field}
\label{sec:voidsinhalo}

The left panel of \reffig{VSF_halo_particle} shows the void size function of voids identified in the halo field both for the 1-fluid (solid line) and 2-fluid (dashed line) simulations. Based on the fact that no relevant differences have been observed between the halo mass function of the two types of simulations (see figure 3 of \cite{Khoraminezhad:2020zqe}), we are not expecting the void size function to be strongly affected either. We do however notice that the number of small voids identified in the halo field in the 2-fluid simulation is higher than the one in the 1-fluid simulation for all redshifts considered, while for larger voids ($R_v\gtrsim 40\rm{Mpc/h}$) we can see the opposite trend (we identified more large voids in the 1-fluid simulation rather than the 2-fluid one). Nevertheless, these differences are relatively small and almost remain inside the errorbars (which shows the error on the mean obtained from the 8 different realizations). This can be seen more directly in the lower left panel of \reffig{VSF_halo_particle} that shows the ratio of the void size function in the 2-fluid and 1-fluid simulations. We see the most significant difference between 1-fluid and 2-fluid simulations for small voids at $z=3$, where we observe more small voids in 2-fluid simulations with a significance of roughly $1.5\sigma$. We see the opposite effect for larger voids but with larger errorbars and consistent with 1. We emphasize the fact that the observed trend is something that we are expecting, since clustering is slightly diminished in 2-fluid simulations. Indeed, in \cite{Khoraminezhad:2020zqe}, figure 9, we have shown that the amplitude of the ratio of the halo-halo power spectrum in 2-fluid simulation over the 1-fluid case is below 1, confirming the fact that baryon-photon coupling in the early Universe decreases the clustering in 2-fluid simulations. Hence, we expect to have more small voids and less large voids in 2-fluid simulations, and we expect this effect to be more important at higher redshift since gravitational evolution washes out relative baryon-CDM perturbations after decoupling. We also note that the effect of baryon-CDM perturbations on the cumulative VSF is smaller than the effect caused by massive neutrinos (see for instance figure 2 of \cite{Massara:2015} in which the authors observed an impact due to neutrino masses up to $\sim 30 \%$ for $\sum m_{\nu}=0.6\,\rm{eV}$ at $z=0$). Finally, the left panel of \reffig{VSF_halo_particle} shows that in both types of simulations, ZOBOV found more small voids at lower redshift and more large voids at higher redshift as can be seen in the redshift trend shown by the color bar. This is also something that we expect, as discussed at the end of \refsec{voidfind}.

\subsection{VSF in the particle field}
\label{sec:voidsinpart}

The right panel of \reffig{VSF_halo_particle} presents the VSF for voids found in the particle field. While we found more large voids and less small voids with increasing redshift in the case of halo field voids, here we see that we find more small voids at higher redshift (and symmetrically less large voids). The redshift trend, in this case, is hence different than for halo field voids for which we recall that the average density of tracers in the box is evolving with redshift which is not the case for particles. This confirms, as was shown in various previous works, that the void population depends on the tracer type one is using, in particular on the tracer density and tracer bias (see for example \cite{Contarini2019,Sutter2014}). The particle field voids are smaller and found in greater numbers than the voids in the halo field. This is due to the fact that the distribution of collapsed halos is sparser than that of cold dark matter particles. These results are again expected, as we discussed at the end of \refsec{voidfind}.

For particle field voids, the difference in the number of voids found in the 1-fluid and 2-fluid simulations is even reduced compared to the halo field void case, and we do not observe any redshift evolution trend of the effect on these VSF caused by the 2-fluid formalism. Hence baryon-CDM relative perturbations impact the VSF of voids identified in halos more importantly, which suggests that they might also impact the VSF of voids found using luminous tracers (such as galaxies) in observations. 


\section{Voids 2-point statistics}
\label{sec:void2pcf}

We now move on to the 2-point statistics of voids, focusing first on the full correlation functions before analysing the voids profile and voids bias in more details.

\subsection{Full correlation functions}
\label{sec:voidfull}

\begin{figure*}
\centering
\includegraphics[width=8.5cm]{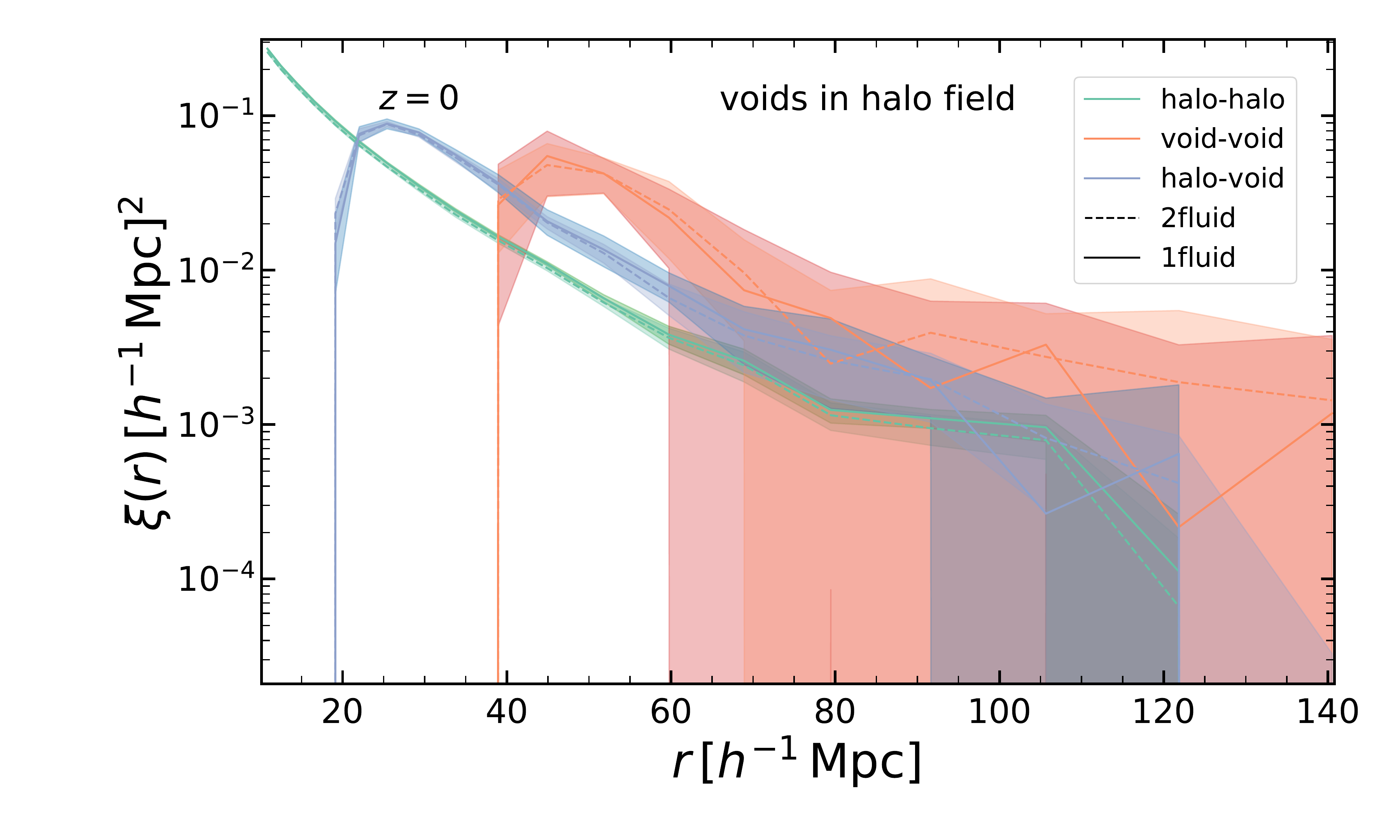}\hfill
\includegraphics[width=8.5cm]{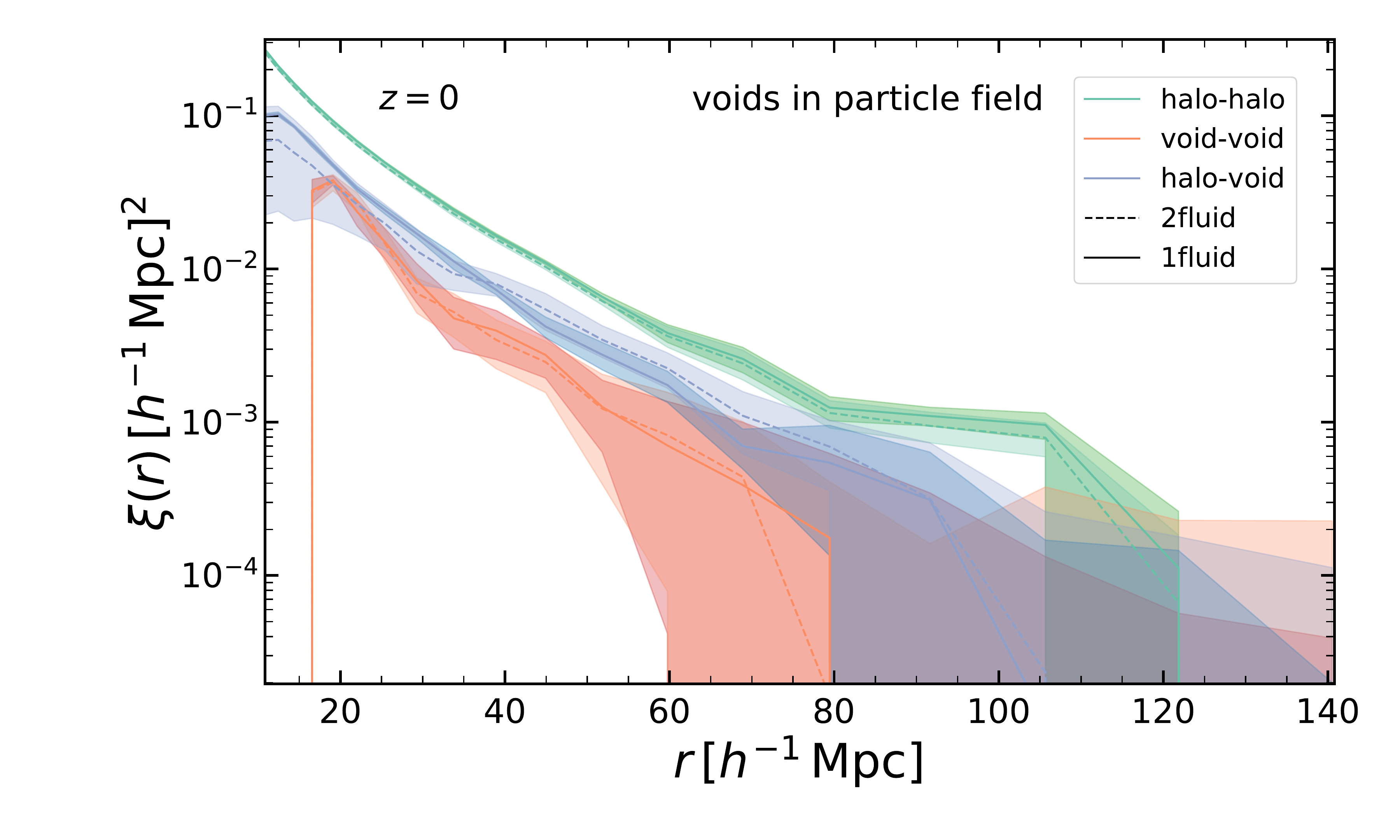}\hfill
\caption{Void-void (orange), halo-halo (green) and halo-void (blue) 2PCF as a function of separation $r$ using voids found in the \textbf{halo field} (left panel) and the \textbf{particle field} (right panel) at $z=0$. The results for the 1- and 2-fluid cases are shown by solid and dashed lines respectively. The shaded area in each case shows the $1\sigma$ error obtained from $8$ realizations.}
\label{fig:void-halo-halo-particle-field}
\end{figure*}

The 2-point correlation function (2PCF) of a set of objects is a measurement of the degree of clustering of the considered objects defined as the excess probability of finding an object at a given distance from another one with respect to a homogeneous distribution of objects. Estimators of the 2PCF, $\xi(r)$, in which $r$ denotes the comoving separation, have been studied by various authors (see for example \cite{peebles_1974,Davis_peebles_1983,Hewett_1982,Hamilton_1993,Landy1994}). Among them, we use the so-called ``natural'' estimator (\cite{peebles_1974}) which has been implemented in the nbodykit pipeline\footnote{https://github.com/bccp/nbodykit} (\cite{Hand2018}) to measure the void and halo auto/cross 2PCF in our simulation boxes.
\begin{equation}
\xi(r)=\frac{DD(r)}{RR(r)}-1 ,
\label{eq:natural}
\end{equation}
which calculates the 2PCF using a data catalogue D, and a synthetic random catalogue R. $DD(r)$ and $RR(r)$ represent the pair counts with separation $r$ in the data and random catalogues respectively. Notice that, in order to reduce computational cost, nbodykit analytically estimates the random pairs $RR(r)$ in the case of uniform periodic randoms such as for simulated boxes.

\refFig{void-halo-halo-particle-field} shows the void-void, halo-void and halo-halo auto(cross)-correlation functions at $z=0$ in 1- and 2-fluid simulations (solid and dashed lines respectively), for voids identified both in the halo field (left panel) and in the CDM particle field (right panel). These correlations are computed for all halos and voids without binning in size. For all cases, the 2PCF is monotonically decreasing as a function of distance. In both panels, the amplitude of the halo-void correlation function stands between the halo-halo and void-void ones for all separations $r$. The halo-halo correlation function (green curve) is the same in both panels (since it obviously does not depend on the tracer used to find voids), and serves as reference to compare the two cases. In the case of voids in the halo field, the amplitude of the halo-void and void-void cross/auto correlations is higher than the halo-halo case, while in the particle field, the halo-void and void-void 2PCFs are lower than the halo-halo one. This demonstrates that, as expected, voids identified in the halo field are more correlated with halos than the voids found in the particle field. Another important feature here is that since voids are larger in the halo field than in the CDM particle field, the void-void 2PCF (orange curve) in the left panel starts to be nonzero at larger separation than the one in the right panel due to the exclusion effect. Indeed, since voids are low-density regions extending several tens of megaparsecs (hence with little amount of tracers inside them), the signal at scales inside the void radius becomes really low (or even zero) when computing the correlation function (or power spectrum) due to the lack of objects inside the voids, (see for instance \cite{Hamaus:2013qja,Chan:2014qka,Platen:2007ng}). This also has for effect to increase the amplitude of the correlation on larger scales in the halo field since larger voids (corresponding to a merging of small ones) can form in the halo field. Finally, we further note that due to the much larger number of halos in comparison to voids ($\sim$ 150 times larger) the signal to noise is much higher for the cross correlation than the auto correlation of voids. This for instance will have a consequence on the precision of the void bias estimation (see \refsec{voidbias}). 

We now inspect in more details the impact of baryon-CDM relative perturbation on the 2PCFs by comparing results in the 1- and 2-fluid cases (solid versus dashed lines). We see that all differences are very small and well within $1\sigma$ errorbars. The largest difference is seen in the case of the halo-void correlation function for voids identified in the particle field (blue lines in the right panel), with the 2PCF computed in the 2-fluid case being slightly smaller at small scales and slightly larger at larger scales. Moreover, we see a small trend on the halo-halo 2PCF, where the 2PCF computed using 2-fluid simulations seems always slightly below the one computed from 1-fluid simulations. This suggests that baryon-CDM relative perturbations tend to lower the clustering, which is in agreement with the expectation of baryon-photon coupling slowing down the clustering process (as discussed in \cite{Khoraminezhad:2020zqe}). However, this effect is quite small and still within our errorbars. Note that this effect is also in agreement with the one we mentioned in \refsec{voidsinhalo} for the VSF, regarding the fact that since we have less clustering in 2-fluid simulations we identify more small voids and less large ones. 

\subsection{Density profiles}
\label{sec:voidprofile}

Cosmic voids are underdense regions close to their center with an overdense compensation wall at $r \sim 2 R_v$, $r$ being the radial distance from the center of the void. Moreover, the deepness of the void center, as well as the amplitude of the compensation wall have been shown to strongly depend on the void population considered (see for example \cite{PhysRevLett.112.251302,10.1093/mnras/stt1069,10.1093/mnras/stu307}). The density profile of voids encodes the same information as the void-tracer cross correlation function since the radial profile of voids is indeed equal to the way that we count the number of tracers at distance $r$ from the center of the void (see \cite{Hamaus:2015yza,Pollina:2016gsi} for a detailed explanation). In more details, the average radial number density of tracers at distance $r$ from the void center, $\rho_{vt}(r)$ (also known in the literature as the \textit{void stacked profile}), can be written as 
\begin{align}
\dfrac{\rho_{vt}(r)}{\langle \rho_t \rangle} &= \dfrac{1}{N_v} \sum_{i} \dfrac{\rho^i_{vt}(r)}{\langle \rho_t \rangle} \nonumber \\
&= \dfrac{1}{N_v}\sum_i \dfrac{1}{N_t} V\sum_j \delta^{D} (x_i^{\rm {center}} - x^t_j +r) \nonumber \\
&= V \sum_{i,j} \int \dfrac{1}{N_v} \delta^{D}(x_i^{\rm {center}}-x) \dfrac{1}{N_t} \delta^{D}(x-x^t_j +r) d^3 x \nonumber \\
&= \dfrac{1}{V} \int \dfrac{\rho_v(x)}{\langle \rho_v \rangle }  \dfrac{\rho_t(x+r)}{\langle \rho_t \rangle } d^3 x = 1+\xi_{vt}(r),
\label{eq:stacked-void}
\end{align}
where $N_v$ and $N_t$ are the number of voids and tracers respectively (with $\langle \rho_v \rangle$ and $\langle \rho_t \rangle$ their respective mean density), $V$ is the total observed volume, $x$ denotes the position (we use the index $i$ to run over voids and $j$ to run over tracers), and $\delta^D$ is the Dirac delta function. We have used the definition of the density of tracers within the void as a sum of Dirac deltas in the second equality, which can then be written as a convolution of the number density of the center of the voids $\rho_v$ and the number density of tracers $\rho_t$ (third and fourth equality), which is finally the definition of the void-tracer cross correlation function $\xi_{vt}(r)$. 

\begin{figure*}
\centering
\includegraphics[width=8.5cm]{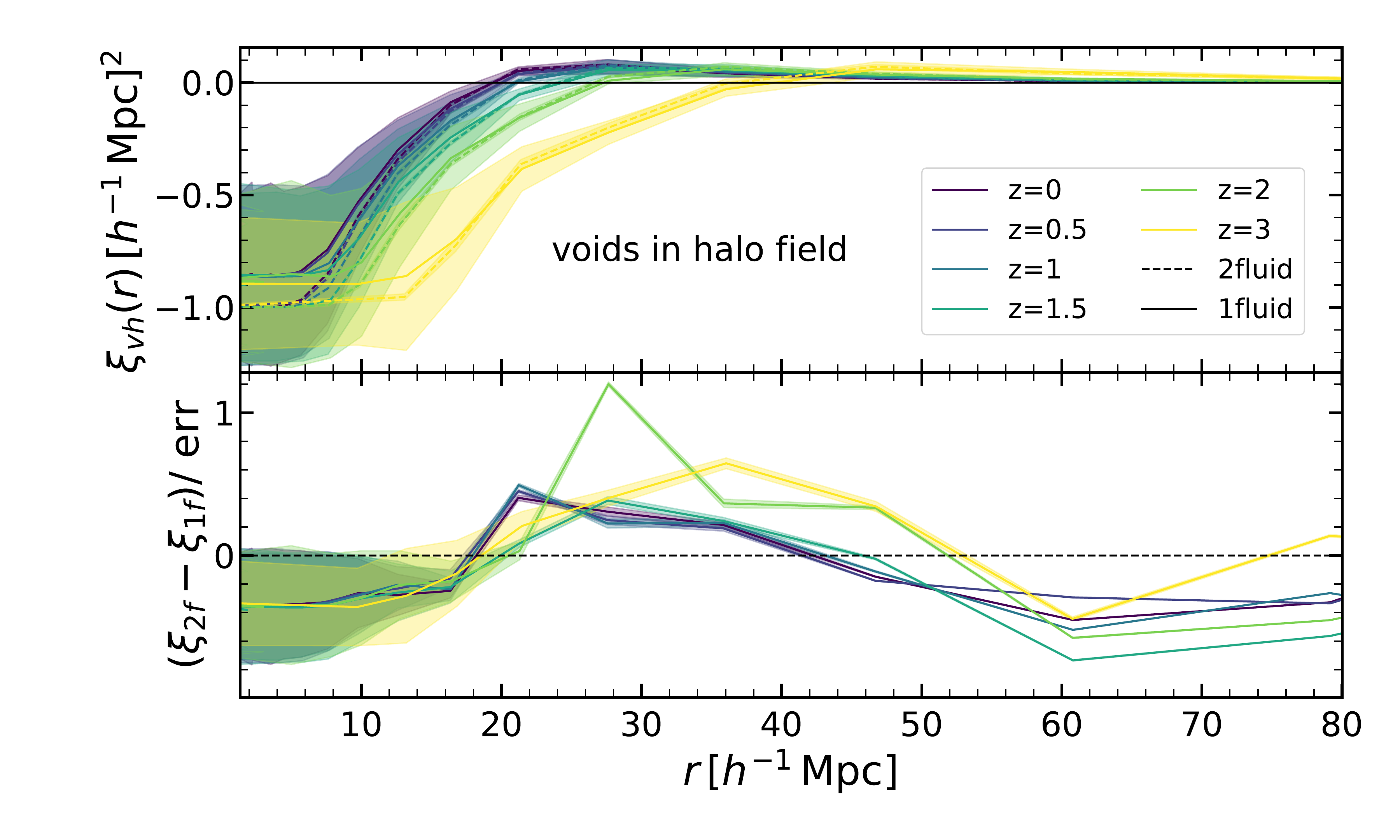}
\includegraphics[width=8.5cm]{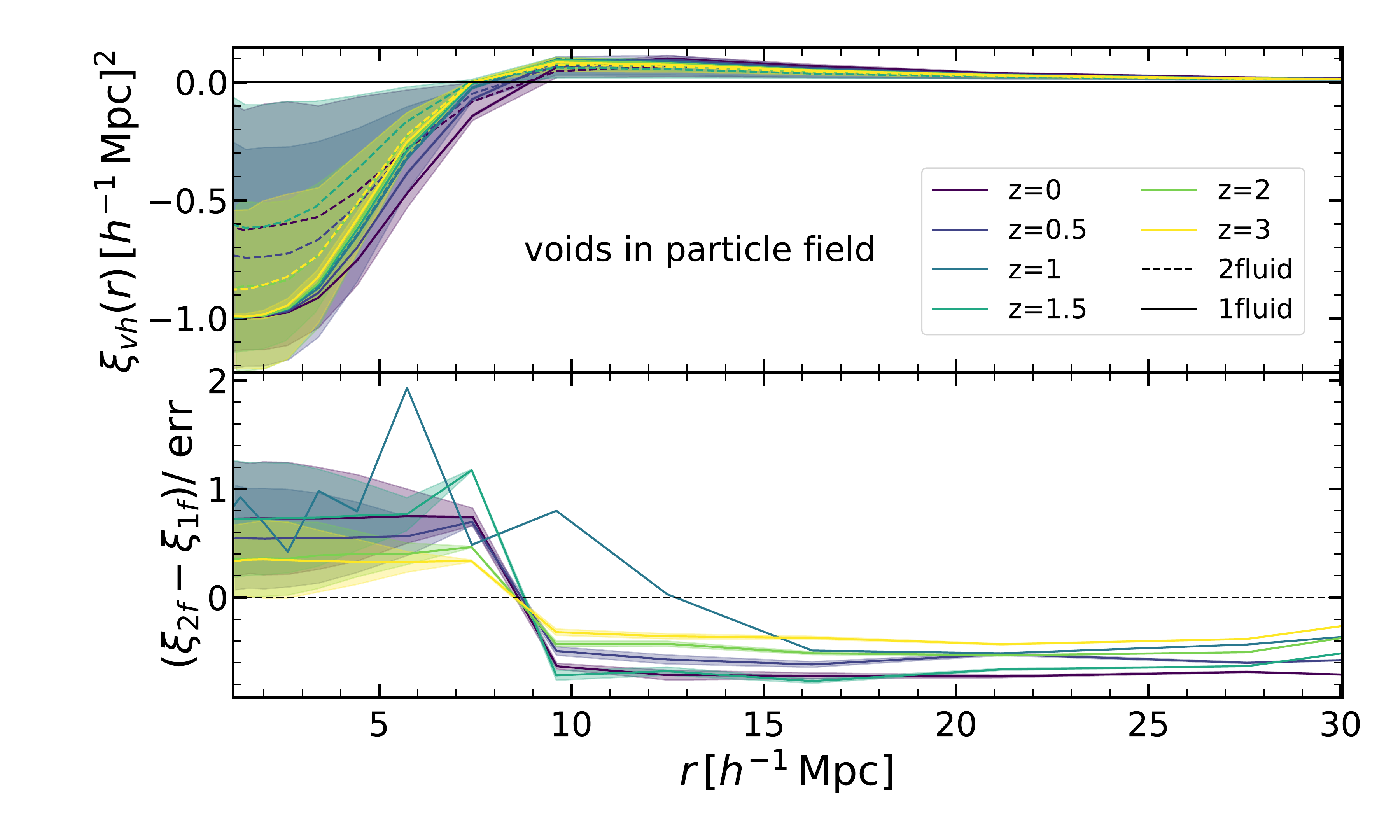}
\caption{Halo-void cross-correlation corresponding to the void stacked profile for voids in the \textbf{halo field} (left panel) and in the \textbf{particle field} (right panel) at different redshift, and for 1-fluid and 2-fluid simulations (solid and dashed lines). We computed $\xi_{vh}$ for all voids in our catalogues (i.e. without applying any cut in radius). Lower panels show the difference between 1-fluid and 2-fluid simulations over the parameter ``$\rm{err}$'', defined as $\rm{err}=\sqrt{\rm{err}_{\xi_{2f}}^2 + \rm{err}_{\xi_{1f}^2}}$.  Note that the curves at $z=0$ are equivalent to the blue curves in \reffig{void-halo-halo-particle-field}, with a vertical axis in linear scale.}
\label{fig:void-profile_all}
\end{figure*}

We use this definition and compute the mean void profile as the halo-void cross correlation function for voids identified both in the halo and particle field. The void density profile for different redshift and different simulations scenarios (1-fluid and 2-fluid) are presented in \reffig{void-profile_all}. The left and right panels display the density profiles of the voids identified in halo and particle fields respectively. Note that \reffig{void-profile_all} is similar to the blue curve in \reffig{void-halo-halo-particle-field} but with a linear vertical axis, and for different redshift represented by the color bar. In \reffig{void-profile_all}, we can distinguish 3 different scales with 3 different behaviors in the density profile: 
\begin{enumerate}
    \item The innermost scales $(\sim r<\bar{R}_{v} /2)$ ($\bar{R}_{v}$ is the mean void radius) in which $\xi_{vh}$ approximately tends to $-1$ at the void centers. Note that since the central part of voids is not totally empty, the cross correlation is not exactly equal to $-1$. 
    \item The intermediate scales $(\sim \bar{R}_{v} /2<r<2\bar{R}_{v})$ or the void profile regime, on which we can see the compensation wall of the voids, which is a positive correlation around the void at all redshift. Notice that for voids identified in the halo field (left panel) the compensation wall moves to higher scales with increasing redshift. This is caused by the fact that the VSF at higher redshift is shifting towards larger radius voids (see \reffig{VSF_halo_particle}, left panel). On the contrary, in the case of particle field voids (right panel), we see that the compensation wall moves towards lower scales with increasing redshift, which corresponds to the fact that the VSF of particle field voids at higher redshifts is shifting towards smaller radius voids (\reffig{VSF_halo_particle}, right panel).
    \item The linear regime $(\sim r > 2\bar{R}_{v})$ in which we see that the compensation wall disappears and $\xi_{vh}\rightarrow 0$. This is the regime in which we will compute the void bias in \refsec{voidbias}. 
\end{enumerate}
Comparing the left and right panels of \reffig{void-profile_all}, we can also see that halo field voids have a much larger mean size than that of the particle field ones. This behaviour is confirmed by the VSF in \reffig{VSF_halo_particle}. The bottom panels of \reffig{void-profile_all} present the difference between 2-fluid simulations and the 1-fluid case over the error parameter which describes the quadrature summation of the errors in each case.  We see that for the halo field voids, for small scales that are inside the void radius, the difference between the 2-fluid and 1-fluid correlation functions is slightly lower than zero at all redshifts, suggesting that 1-fluid voids are somewhat smoother (recall that the density is negative on those scales). This effect is within errorbars but can be seen for the mean value of the difference for halo field voids. Moreover, this effect is not seen in particle field voids (right panel) due to the fact that the signal is more noisy since we correlate particle field voids with halos. Finally, we note that errorbars in the void center are quite large due to the low-density definition of voids, and thus the lower amount of halos to compute the correlation.

\begin{figure*}
\centering
\includegraphics[scale=0.4]{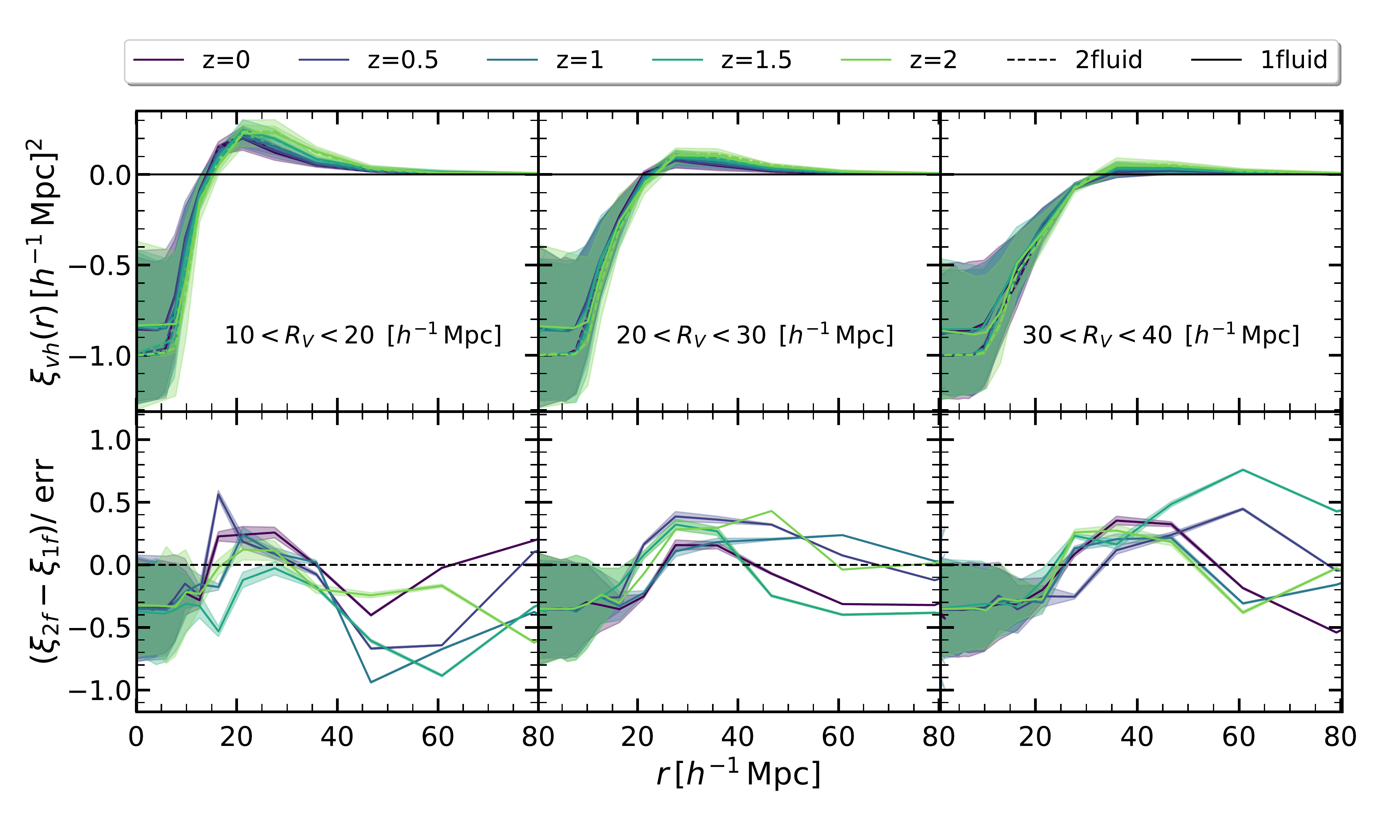}
\caption{Void profile of \textbf{halo field voids} for 3 different bins of void radius $R_v$ for 1-fluid (in solid line) and 2-fluid (in dashed line) simulations at 5 different redshift (color coded). In each bin and at all redshift the cross correlation approaches $-1$ close to the center of the void ($\sim(r<R_{v} /2)$). On scales $\sim(R_{v} /2<r<2R_{v})$, the void profile shows a prominent compensatory ridge of halos for smaller voids $10 < R_v < 20 \,\,\Mpch$, which disappears for the largest voids $30 < R_v < 40 \,\,\Mpch$. In each bin, this compensation wall moves to lower radius (smaller voids) with decreasing redshift, which is the same behavior as we noticed in \reffig{void-profile_all}. Lower panels show the difference between the 1-fluid and 2-fluid case over the error defined as $\rm{err}=\sqrt{\rm{err}_{\xi_{2f}}^2 + \rm{err}_{\xi_{1f}^2}}$ for each void size bin,  where we see that voids in the 2-fluid case tend to be slightly less dense in their center.}
\label{fig:void-profile_bins}
\end{figure*}

\begin{figure*}
\centering
\includegraphics[scale=0.4]{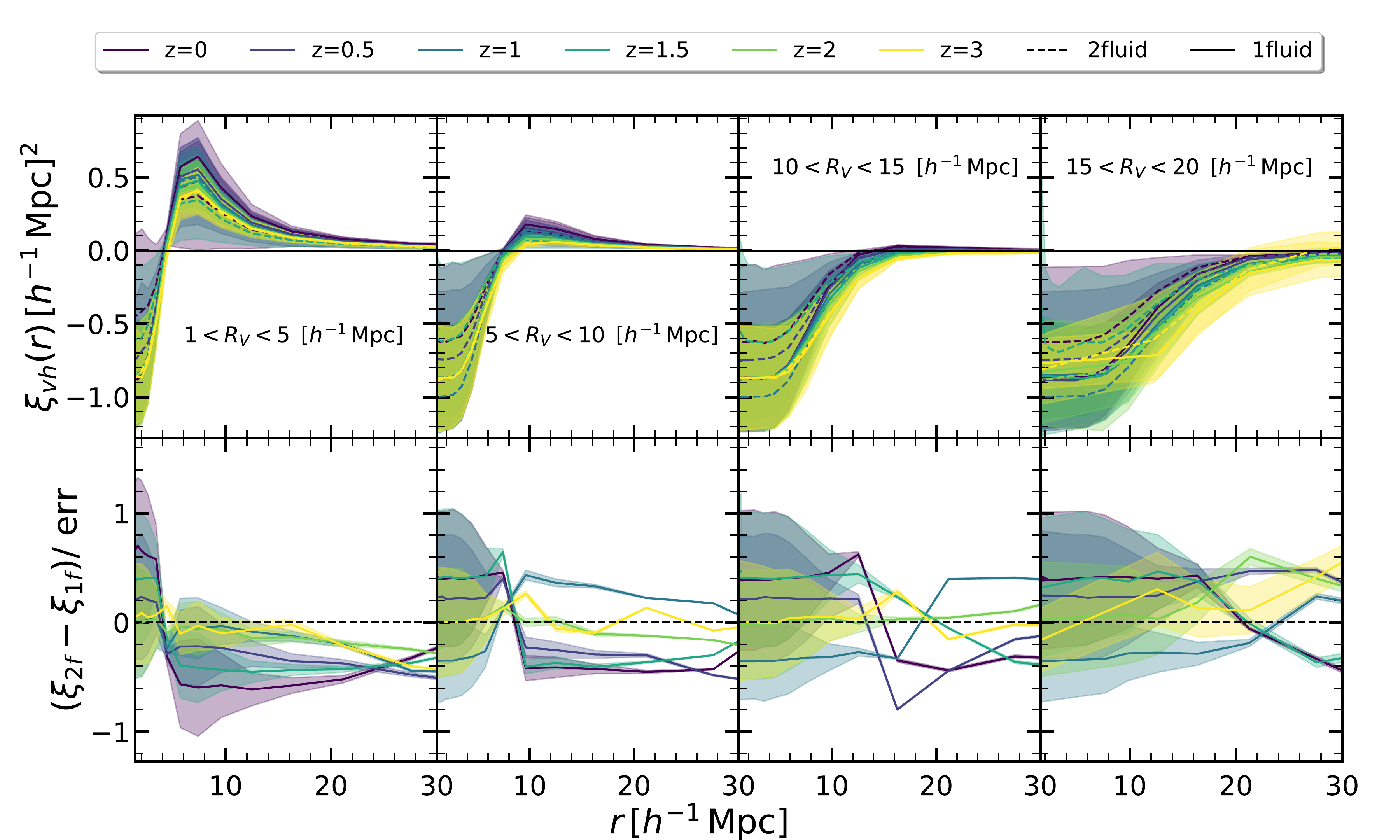}
\caption{Same as \reffig{void-profile_bins} but for \textbf{voids in the particle field} in 4 different bins of void radius $R_v$. Since the number of voids identified in the particle field is larger than in the halo field, we divided them into more radius bins than halo field voids. The results of the 1-fluid scenario are shown by the solid lines and the ones of the 2-fluid by the dashed lines. The color bar represent different redshift. The void profile shows a sizable compensation wall for the voids in the smallest size bin  ($1<R_v<5 \, \Mpch$). When moving to larger voids this structure becomes less prominent. The lower panels show the difference between the 1-fluid and 2-fluid case over the error defined as $\rm{err}=\sqrt{\rm{err}_{\xi_{2f}}^2 + \rm{err}_{\xi_{1f}^2}}$ for each void size bin.}
\label{fig:void-profile_bins_particle}
\end{figure*}

The density profile of voids has been shown to depend on the void size (see for example \cite{PhysRevLett.112.251302,Sutter2014}), and we next explore whether or not the effect due to baryon-CDM perturbations could also vary with voids size. We divided our catalogues of void identified in the halo field in 3 different radius bins: $10 < R_v < 20 \,\,\Mpch, \,\, 20 < R_v < 30 \,\,\Mpch, \,\,$, $30 < R_v < 40 \,\,\Mpch$, and the catalogues of void identified in the particle field in 4 radius bins: $1<R_v<5\,\, \Mpch$, $5<R_v<10\,\, \Mpch$, $10<R_v<15\,\, \Mpch$, $15<R_v<20\,\, \Mpch$. The void profile (i.e. the void-halo cross correlation function) for each radius bin for each type of voids and at different redshift are shown in \reffig{void-profile_bins} (for halo field voids) and \reffig{void-profile_bins_particle} (for particle field voids). In \reffig{void-profile_bins}, we do not show results at $z=3$ since the number of voids is quite small and the cross correlation signal becomes too noisy. For both types of voids and for all different void size bins, we observe the same 3 different regimes mentioned above (the innermost scale, the intermediate scale and the linear regime). We note that for all types of voids (found in halos or particles) the compensation wall found at intermediate scales (the void profile regime) is more pronounced at smaller radius: in \reffig{void-profile_bins}, we see a clear positive bump in the first panel for smallest halo field voids, and as we move to the second and third panels (to larger voids), the bump becomes less prominent and it disappears in the last panel for the largest voids. We observe the same behaviour in \reffig{void-profile_bins_particle} for particle field voids. The results found here are qualitatively in agreement with \cite{Ceccarelli:2013rza,PhysRevLett.112.251302} and \cite{Clampitt:2015jra}.

In the same manner as for the void profiles of all voids (without classifying them by their radius), we show the difference of the results from the 2-fluid scenario and the 1-fluid one over the quadrature summation of the errors in each case in the lower panels of \reffigs{void-profile_bins}{void-profile_bins_particle}. For halo field voids (\reffig{void-profile_bins}), inside the voids, we see that $\xi_{vh,\rm{2f}} < \xi_{vh,\rm{1f}}$ at all redshift which tells us again that 1-fluid voids are slightly smoother. We do not observe this for particle field voids (\reffig{void-profile_bins_particle}), because the signal is much more noisy again. We note that the effect of baryon-CDM perturbations on void profiles does not seem to depend on the void radius as we observe that the difference seems to be similar inside the voids in all panels. Finally, we emphasis that these differences are always compatible with 1 within $1\sigma$ errorbars, therefore we conclude that there are no significant differences between void profiles in 1-fluid and 2-fluid simulations, and hence that baryon-CDM relative perturbations due to photon pressure do not significantly impact this quantity. The results in each radius bin in \reffig{void-profile_bins} and \reffig{void-profile_bins_particle} are compatible with those obtained for all voids without binning in size (\reffig{void-profile_all}).

\subsection{Void bias}
\label{sec:voidbias}

In addition to the different void observables presented above, we also aim in this work to quantify the impact of baryon-CDM perturbations on the linear bias of cosmic voids. Indeed, the estimation of the clustering bias of cosmic voids is an essential element to achieve competitive cosmological inference from voids, in the same way as galaxy bias in the case of galaxies (\cite{Desjacques:2016bnm,Schmidt:2020tao,Pezzotta:2021vfn} and references therein). In this perspective, the interest in understanding it and modeling is raising (\cite{Sheth2004,Hamaus:2013qja,Chan:2014qka}). Moreover, the possibility of using void bias directly to constrain cosmology is also recently gaining interest (see for example \cite{Schuster2019,chan2019,chan2020}).  
Here, we will measure the bias of our voids following the methodology described in \cite{Clampitt:2015jra}, for both 1-fluid and 2-fluid simulations at different redshift, and considering both voids identified in the halo field and in the particle field. Similarly to \cite{Clampitt:2015jra}, we define the void bias using two different expressions, the first one using the halo-void cross-correlation as
\be
b_v^{\textrm{cross}} = \dfrac{\xi_{vh}}{b_h \, \xi_{mm}},
\label{eq:bias-cross}
\ee
in which the halo bias can be obtain using the halo auto-correlation signal as $b_h \equiv \sqrt{\xi_{hh}/\xi_{mm}}$. Thus one can rewrite \refeq{bias-cross} as
\be
b_v^{\textrm{cross}} = \dfrac{\xi_{vh}}{\sqrt{\xi_{hh}\, \xi_{mm}}}.
\label{eq:bias-cross-main}
\ee
The second definition uses the void-void auto-correlation as follows
\be
b_v^{\textrm{auto}} = \pm \sqrt{\dfrac{\xi_{vv}}{\xi_{mm}}},
\label{eq:bias-auto-main}
\ee
where in all the above equations $\xi_{mm}$ is the matter-matter auto correlation function measured directly from the simulation snapshots (using only CDM in 2-fluid simulations), and $\xi_{hh}$ is the halo-halo auto correlation function shown in \reffig{void-halo-halo-particle-field}. For $b_v^{\textrm{auto}}$, we first measure the bias squared and then we chose the sign of the square root using the sign of $b_v^{\textrm{cross}}$ (identically to what has been done in \cite{Clampitt:2015jra}). Considering the number of voids in each bin, we expect $b_v^{\textrm{auto}}$ to be much more noisy. Nevertheless, it is interesting to cross-check to see if both bias measurements give comparable values.

\begin{figure*}
\centering
\includegraphics[width=8cm]{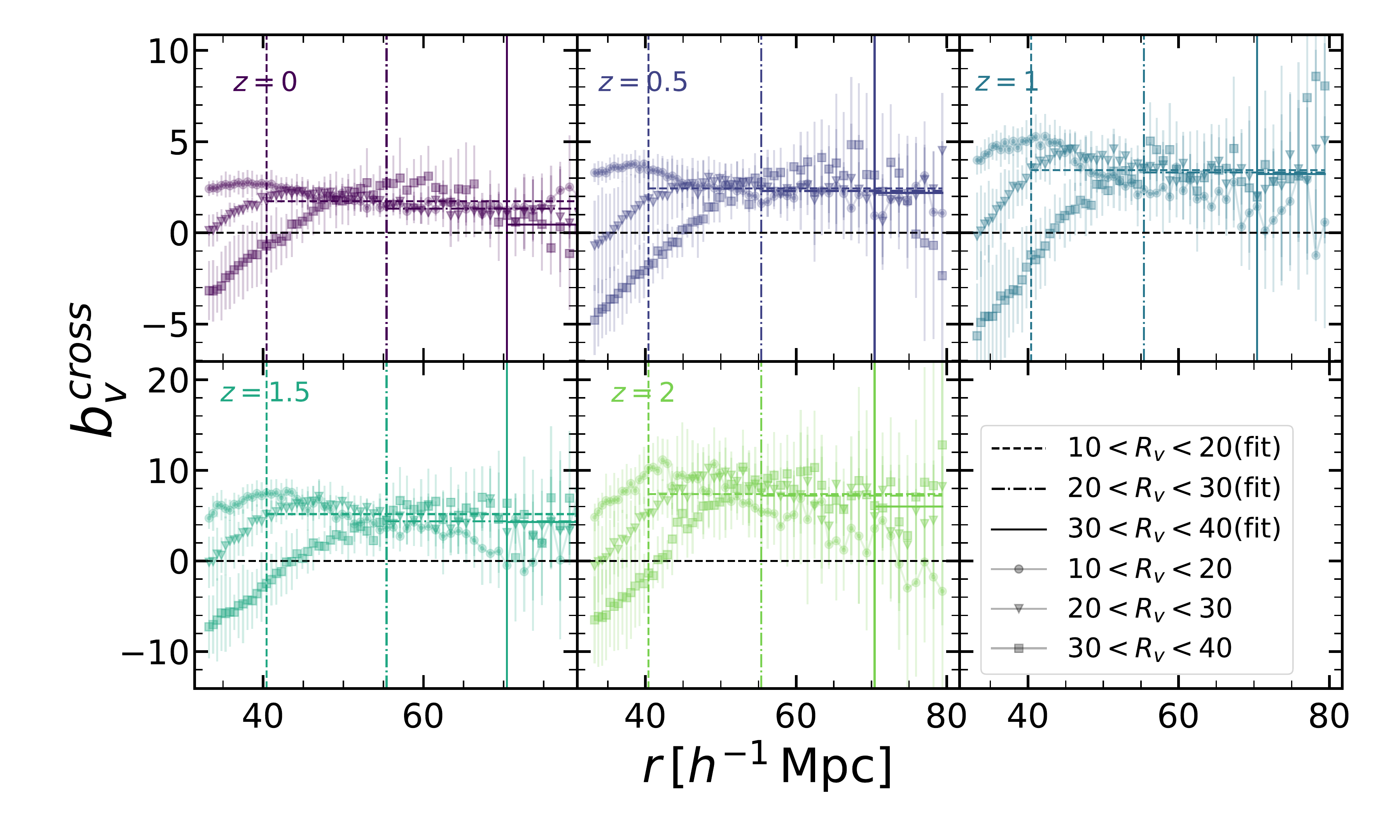}\hfill
\includegraphics[width=8cm]{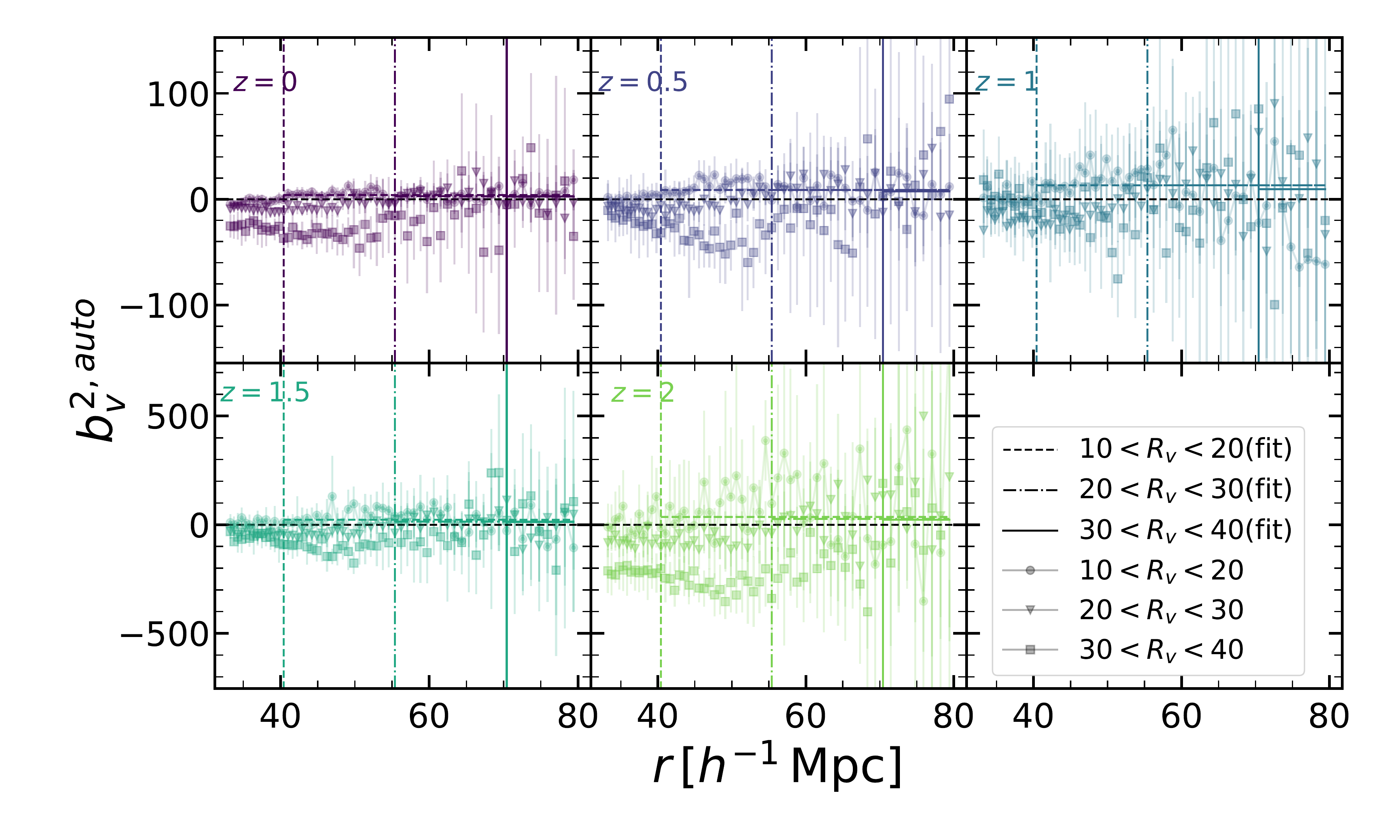}\hfill
\caption{$b_v^{\textrm{cross}}$ (left) and $(b_v^{\textrm{auto}})^2$ (right) as a function of scale (\refeqs{bias-cross-main}{bias-auto-main}) and an example of the fit with a zeroth order polynomial to obtain the mean void bias value. Both panels present results from 2-fluid simulations for voids in the halo field. Each subpanel with different color presents results at a different redshift. Different markers and line styles show the measurement and associated fit at different void radius $R_v$. The vertical line in each line style is showing the $2R_v$ value for each bin. The errorbars show the $1\sigma$ error on the mean obtained from 8 realizations. Since the number of voids is roughly $\sim 150$ times less than the number of halos, we have very large errors when computing $(b_v^{\rm{auto}})^2$.}
\label{fig:fitting}
\end{figure*}

\refFig{fitting} presents, as an example, the cross (left panel) and auto (right panel) bias as a function of scale for halo field voids at various redshift and void radius. Each small panel with different color presents a different redshift. Considering the few number of voids identified at $z=3$ and the low signal to noise ratio resulting, we do not show the bias analysis results at $z=3$. We use different markers for different void size bins. The markers here show the mean value of the bias and the errorbars are the 1$\sigma$ error over 8 realizations. As expected, in the linear regime both bias are showing a constant behaviour. We then obtained the values for $b_v^{\textrm{cross}}$ and $(b_v^{\textrm{auto}})^2$ as a function of redshift and void size by fitting a zeroth order polynomial on linear scales (horizontal lines in the figure). In both cases, we use only scales between $2 R_v<r(\rm{Mpc/h})<80$ for the fit. The lower limit assures that we are using only pairs of distinct voids, and the upper limit assures us to avoid the BAO scale on which dividing by $\xi_{mm}$ would create a high noise. We use different line styles to show the fit in different size bins, and we show here the fit over the mean values taking into account the errorbars over different realizations. We also did the same fit for each of the realization to find the errorbars over the mean value of the bias from 8 realizations. As expected, we observe a higher amount of noise in $(b_v^{\textrm{auto}})^2$ than in $b_v^{\textrm{cross}}$ (notice the difference in $y$-axis range) due to the fact that the pair counts in $\xi_{vv}$ are much smaller than $\xi_{vh}$. In addition, the errorbars are increasing with redshift due to the smaller amount of voids found at higher redshift. Regarding the values of $b_v^{\textrm{cross}}$ and $(b_v^{\textrm{auto}})^2$, since the linear bias of halos is increasing with redshift (e.g. \cite{Tinker:2010my}), one can expect the voids identified with this tracer to also become less biased as time evolves, which is indeed what we observe. We also see that the void bias slightly decreases with increasing void size which is in agreement with the results in \cite{Clampitt:2015jra,Hamaus:2013qja}. 

\begin{figure*}
\centering
\includegraphics[width=8cm]{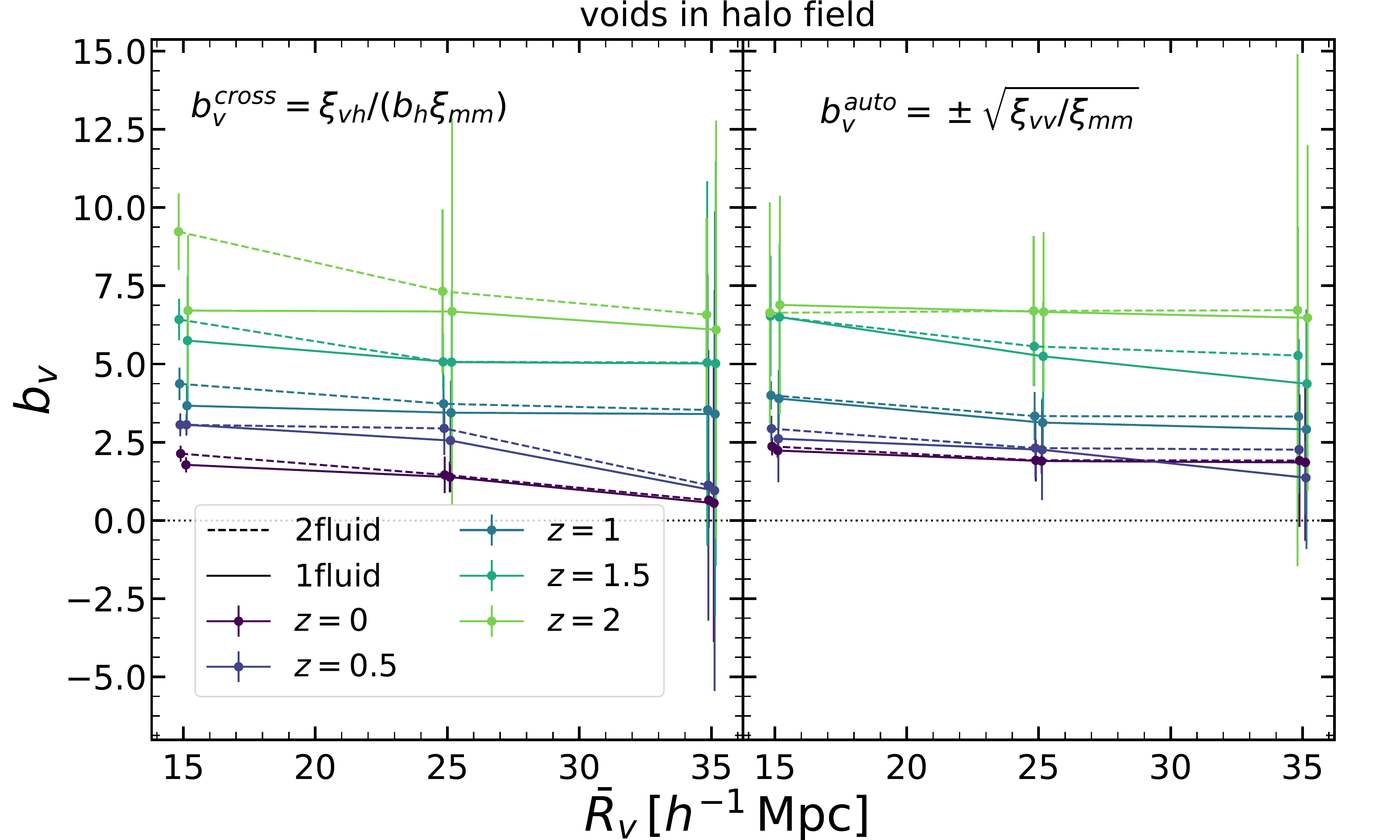}\hfill
\includegraphics[width=8cm]{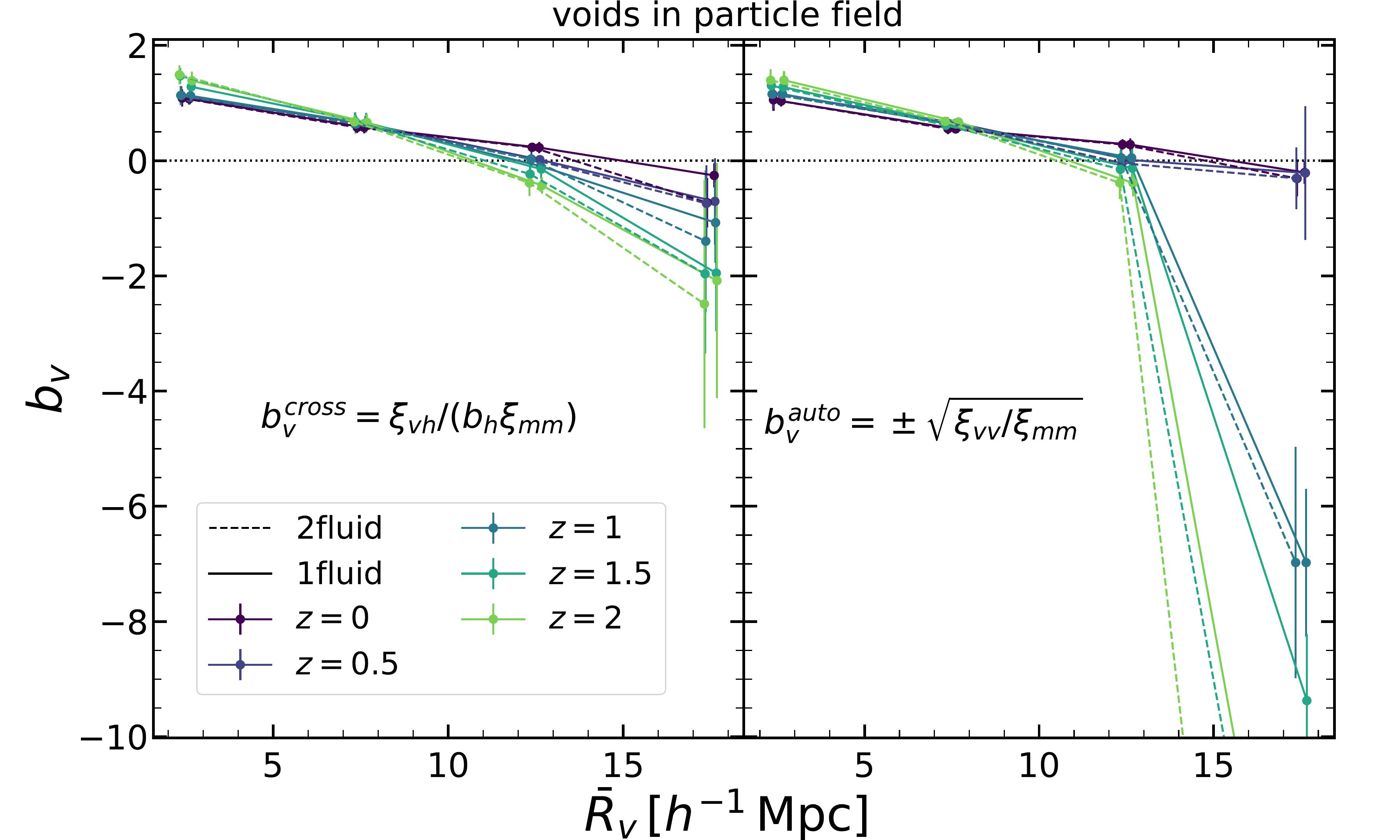}\hfill
\caption{Mean, scale-independent void bias as a function of mean void radius $\bar{R}_v$ obtained from the fits on \reffig{fitting}. \textbf{Left panels:} $b_v^{\textrm{cross}}$ and $b_v^{\textrm{auto}}$ obtained using voids identified in the halo field for all different redshift (color coded), for both 1-fluid (solid lines) and 2-fluid (dashed lines) simulations. We see that both $b_v^{\textrm{cross}}$ and $b_v^{\textrm{auto}}$ slightly decrease with increasing void size, and that both increase with increasing redshift. \textbf{Right panels:} same as the left panels but for voids found in the CDM particle field. In this case, we see that both biases depend more strongly on the void radius, and larger voids become negatively biased at all $z$. We further see that all voids become more positively bias \textit{and} more negatively biased with increasing redshift. We observe only small differences that are all within 1$\sigma$ errorbars between the void bias measured from 1- and 2-fluid simulations.}
\label{fig:void-bias-halo}
\end{figure*}

We then show in \reffig{void-bias-halo} the mean void bias as a function of the void radius integrated over the scales mentioned above (i.e. the value of the fits obtained on scales $2R_v < r <80 \Mpch$). We show both void bias results from cross-correlation, $b_v^{\textrm{cross}}$, and auto-correlation $b_v^{\textrm{auto}}$ using different tracers (in left panels we present results in the halo field and the right panels show results in the particle field). Different colors show different redshift as before. Since in the particle field we have a much larger number of voids, the errorbars are quite smaller compared to the halo field results. Moreover, the number of voids in both halo and matter fields drops significantly with increasing redshift, resulting in more noisy correlation measurement, and consequently, the errorbars of our void bias measurement are also increasing with redshift. This is the main reason why we do not show results at $z=3$. 

Inspecting \reffig{void-bias-halo} in more details, we see that measurements of the bias from the two definitions, i.e. using either the auto (\refeq{bias-auto-main}) or the cross (\refeq{bias-cross-main}) correlation signals, are broadly consistent for all void size bins considered, both for halo field and for particle field voids, except for the highest radius bin of the particle field voids. However, this is probably due to the fact that the signal in this case is really noisy due to the low number of objects, which affects our measurements and might lead to a slight underestimation of the errorbars. A detailed investigation is beyond the scope of this work in which we focus on the comparison between 1- and 2-fluid simulations. If we now inspect the difference between halo field and particle field voids, we see that choosing different tracer significantly affects the void bias: voids identified in the halo field are more biased than the particle field voids which is something expected since dark matter halos are biased themselves. Furthermore, we find that in the case of the voids in the halo field, the mean value of the void bias is a slightly decreasing function of the void size (almost consistent with a constant considering the errorbars), while for the particle field, the void bias is a decreasing function as the size of the voids is increasing. In the right panel of \reffig{void-bias-halo}, we observe that the particle field void bias changes sign at a specific ``turning scale'', which is a similar behaviour as observed by \cite{Clampitt:2015jra}, with however a different turning scale. This turning scale is roughly at $\sim 15 \Mpch$ for our voids in the particle field while roughly at $\sim 25 \Mpch$ for SDSS voids in \cite{Clampitt:2015jra}. However, we do not expect to observe the change of sign at the exact same scale since these authors find voids in a different tracer field using a different void finder. 

Comparing the void bias from 1-fluid and 2-fluid simulations (solid versus dashed lines), we see that voids from the 2-fluid simulations are slightly more biased for both voids from the halo field and the particle field. This difference is within 1$\sigma$ errorbars, but the trend of the 2-fluid simulation bias being slightly larger is expected: since the 2-fluid halo-halo 2PCF (the green curves in \reffig{void-halo-halo-particle-field}) is showing less clustering than in the 1-fluid scenario, the linear halo bias $b_h$ is expected to be smaller in 2-fluid simulations as well. Then we can see from \refeqs{bias-cross}{bias-auto-main} that the void bias should be slightly larger in the 2-fluid case. 

\section{Baryon acoustic oscillations}
\label{sec:bao}

In this section, we extend the computation of the real-space 2-point correlation function in 2-fluid simulations from voids to each component of the simulations, i.e. total matter, baryons only, CDM only, baryon-CDM relative perturbations ($\dbc$), and halos. In particular, we focus on modulations of the BAO feature and BAO peak position by comparing our results for the total matter and halo fields in 1- and 2-fluid simulations. This is a direct extension of our previous work (\cite{Khoraminezhad:2020zqe}) where we focused on Fourier space quantities.

Relative velocity perturbations between baryons and CDM can possibly shift the BAO scale because they are sourced by the same physical effect which imprinted the BAO peak itself. The shift in the BAO scale is crucial for cosmology since it could lead to a potential systematic shift in measurements of the angular diameter distance $D_{A}(z)$, the Hubble factor $H(z)$, and the growth factor $f\sigma_8$ (\cite{Dalal_2010,Yoo:2013qla,Beutler:2016zat,Barreira:2019qdl}). This effect might also be important to obtain unbiased results when one is investigating the effect of massive neutrinos on the BAO scale (\cite{Peloso:2015jua}) or when one is using reconstruction methods to measure the BAO location in 21 cm intensity mapping surveys (\cite{Villaescusa-Navarro:2016kbz,Obuljen:2016urm}). 

\subsection{Full-shape correlation function}
\label{sec:fullcorr}

In this subsection, we first focus on the full shape of the 2-point correlation function. To compute the 2-point correlation function in real-space, we use the Fast Fourier Transform (FFT) estimator introduced in \cite{Taruya_2009} in which the density field is computed on a grid in Fourier space, squared, inverse Fourier transformed, and averaged in radial bins  
\be
\xi_{\rm{SIM}} (r) = \dfrac{1}{N_{\rm{modes}}} \sum_{r_{\rm{min}}<|r|<r_{\rm{max}}} \rm{FFT}^{-1} \left[ |\delta(k)|^{2} \right](r), 
\label{eq:2pcf-sim}
\ee
where the sum runs over all radii $r$ in the bin and $N_{\rm modes}$ is the number of modes in the bin. We use the Cloud-In-Cell (CIC) mass-assignment scheme to compute the density field on the Fourier grid $\delta (k)$. To compute the total matter field $\delta_m$ in 2-fluid simulations is given by the weighted sum of the CDM field $\delta_c$, and the baryon field $\delta_b$, as $\delta_m$ as $\delta_m = f_b \delta_b + (1-f_b)\delta_c$, where $f_b= \Omega_b / \Omega_m$. Moreover, we choose the edges of the bins $r_{\rm{min}}$ and $r_{\rm{max}}$ such that each bin as a width given by the mean interparticle separation, which in our case is $512 \Mpch /500 \approx 1\,\, \Mpch$. We take advantage of the fact that this estimator is implemented in the PYLIANS library \footnote{https://github.com/franciscovillaescusa/Pylians}, which we use to obtain our results.  Finally, we restrict ourselves to the real-space 2-point correlation function in 1-fluid and 2-fluid simulations without considering redshift space. The estimator introduced here to calculate the 2PCF is much faster than the natural estimator we were using to compute the void correlation functions and density profiles in \refsec{void2pcf}. There we were using the natural estimator since the sparsity of voids and exclusion effects introduce large noise which prevented us to use the Taruya estimator to obtain the void profile. Here, since we are interested in the correlation function of particles and halos, which are by far more numerous, we can use the FFT estimator to significantly reduce the computation time while keeping a high-level of accuracy. 

\begin{figure*}
\centering
\includegraphics[scale=0.3]{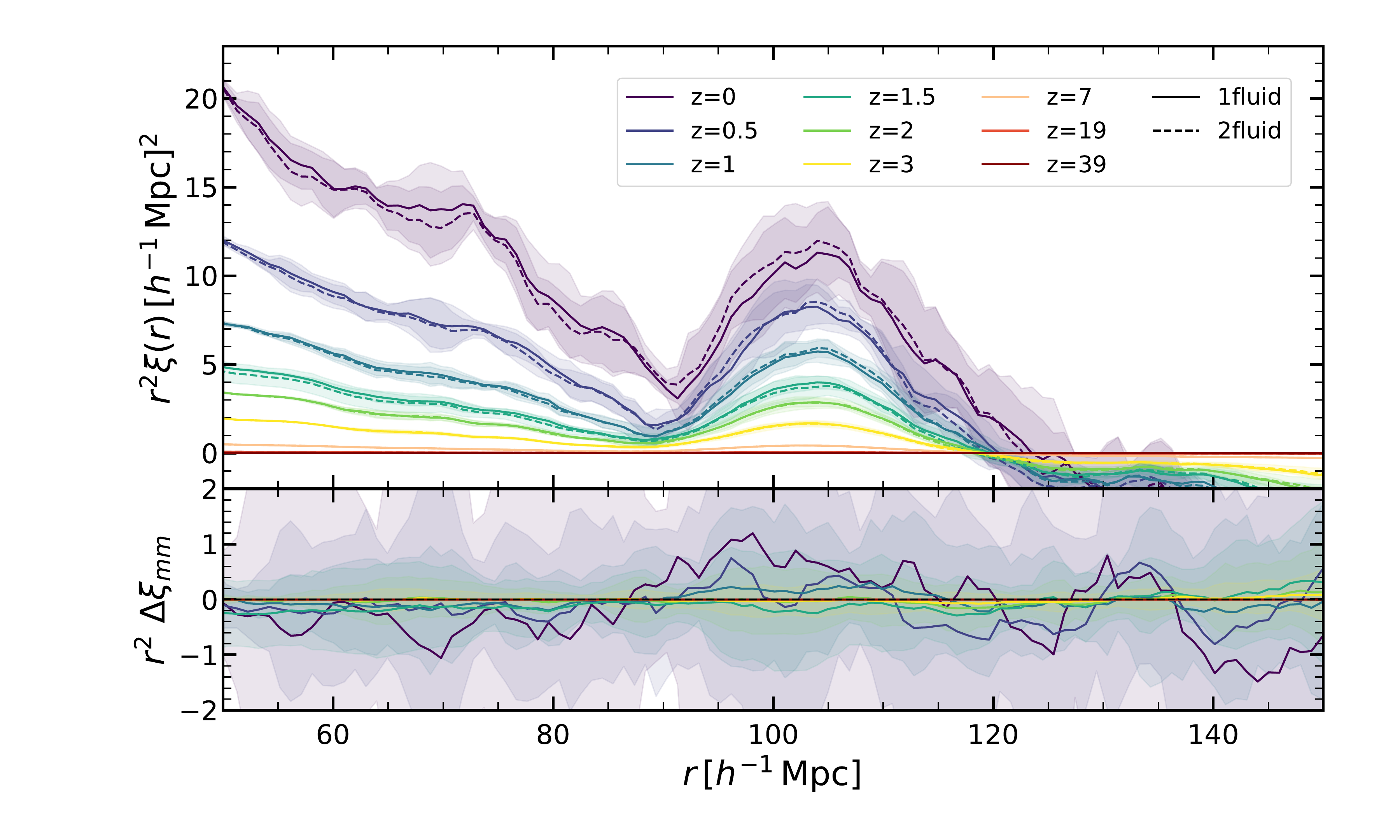}
\caption{The redshift evolution of the full-shape total matter 2-point correlation function in 1-fluid (solid) and 2-fluid (dashed) simulations in real space as measured by \refeq{2pcf-sim}. We multiply the 2PCF by $r^2$ to see the BAO peak better. The shaded area show the $1\sigma$ errorbar on the mean obtained from the standard deviation over all realisations. The lower panel presents the difference between 2-fluid and 1-fluid sets: $r^2\,\,\Delta \xi_{mm}=r^2\,(\xi_{mm}^{\rm{2f}}-\xi_{mm}^{\rm{1f}})$. We see that any small difference between the two cases is within the errorbars on all scales.}
\label{fig:CF-real-zdepend}
\end{figure*}

\refFig{CF-real-zdepend} shows the total matter-matter 2PCF computed in 1- and 2-fluid simulations (solid and dashed lines respectively) using \refeq{2pcf-sim} for different redshift (color coded). We recognize the standard shape of the correlation function which decreases as $r$ increases, as well as the BAO peak at around $r\sim 105 \Mpch$. We also see that both the correlation and the BAO peak increase with decreasing redshift since the clustering becomes more important at lower redshift. We observe small differences between the two cases with the correlation function being slightly lower on smaller scales in 2-fluid simulations, while on the scales of the BAO peak, the 2-fluid simulations give us a higher value of the 2PCF, and the effect is more important at low redshift due to nonlinear evolution (recall that the total matter linear power spectrum is kept constant between 1- and 2-fluid simulations). Notice that these differences are within 1$\sigma$ errorbars obtained over different realizations on all scales. These results confirm that baryon-CDM relative perturbations have a rather small impact on the matter clustering (under the detection threshold corresponding to our simulation volume) as was already pointed out in \cite{Angulo:2013qp,Khoraminezhad:2020zqe}.    

\begin{figure*}
\centering
\includegraphics[scale=0.3]{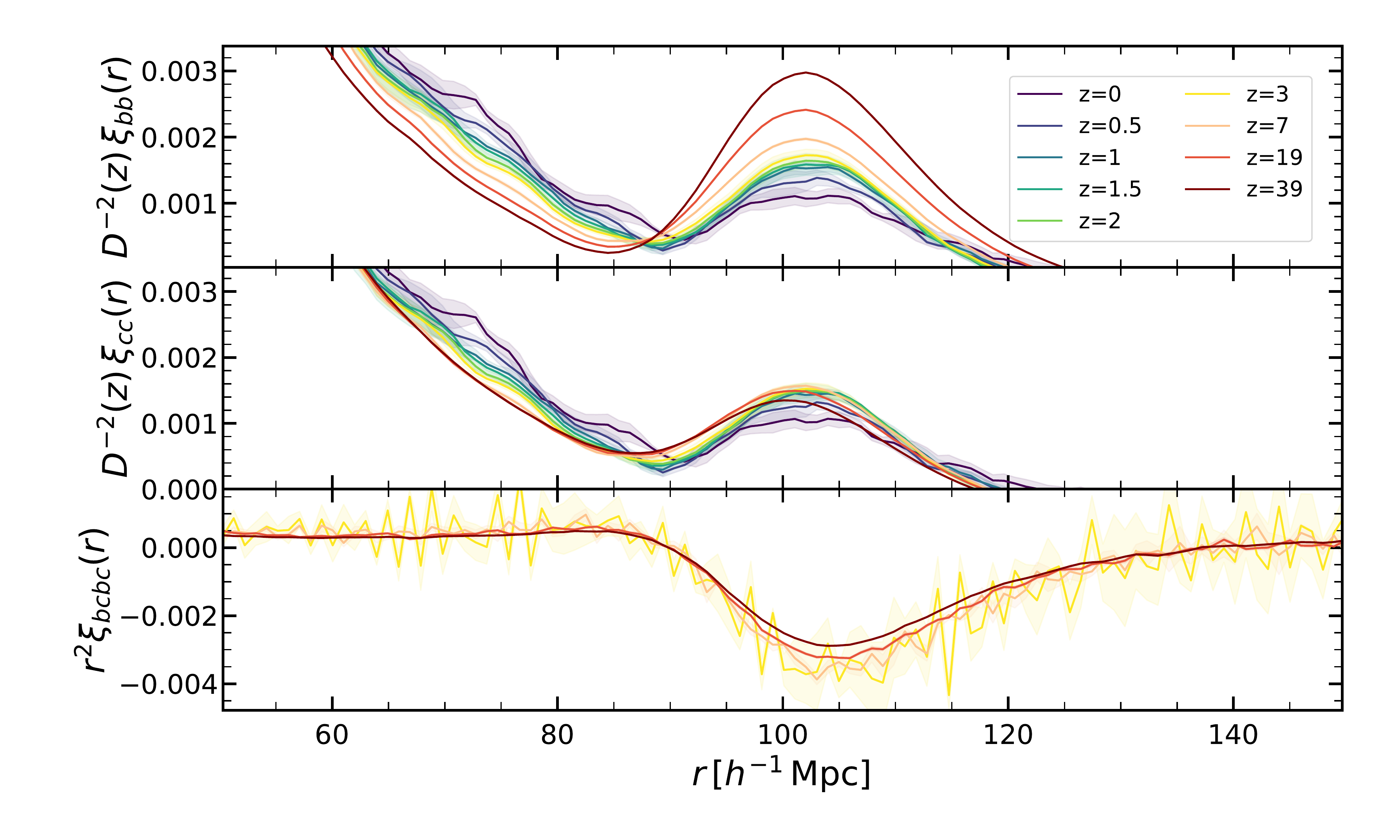}
\caption{\textbf{Top panel:} Baryon 2-point correlation function measured in the 2-fluid simulations at different redshift (color coded). We see clearly how the BAO peak of the baryon 2-point correlation function scaled by $D^{-2}(z)$, diminishes with time in this field. \textbf{Middle panel:} Same as top panel but for the CDM fluid. In this case and in this range of $z$, the BAO peak of the CDM 2-point correlation function scaled by $D^{-2}(z)$, remains roughly constant. We note that the evolution of the baryon-baryon and CDM-CDM correlation functions, \textit{without multiplying by} $D^{-2}(z)$, is the same as what we have shown in \reffig{CF-real-zdepend} which is representing the total matter correlation function. \textbf{Bottom:} The $\delta_{bc}$ relative perturbation auto-correlation function. In this case, we show results down to $z=3$ only since the noise becomes too important at later times. The BAO feature is clearly visible and is negative in this field. Furthermore, we see no redshift evolution, which is consistent with the fact that $\delta_{bc}$ is constant in time, as discussed in e.g. \protect\cite{Schmidt:2016,Hahn:2020lvr,Khoraminezhad:2020zqe}. Note that the two upper panels are divided by the square of the growth factor $D^2(z)$ to see the difference in evolution of BAO in baryons and CDM, while in the bottom panel we multiplied the 2PCF of $\delta_{bc}$ by $r^2$ in order to show the BAO feature better. Shaded area on each curve represent the 1$\sigma$ error, and we see that with increasing redshift the error becomes less prominent.}
\label{fig:CF-bc}
\end{figure*}

We now turn to a more detailed investigation of the cross-correlation of each fluid component in 2-fluid simulations in \reffig{CF-bc}. The top and middle panels show the two different component of the matter field (baryon and CDM) 2PCF divided by the square of the linear growth factor $D^2$. In case of baryons, we can see that the correlation function exhibits a strong  BAO peak at high redshift, and that then the amplitude of the peak decreases with redshift due to gravitational interactions with CDM particles (note that with decreasing redshift the 1$\sigma$ error on the mean value increases). We checked that the evolution of the baryon-baryon and CDM-CDM correlation functions, \textit{without multiplying by} $D^{-2}(z)$, is the same as the total matter one (\reffig{CF-real-zdepend}). Multiplying the baryon-baryon and CDM-CDM correlation functions by $D(z)^{-2}$ effectively removes the linear growth of structure and hence leaves only the fact that the BAO peak decreases with time.  We can also see a small scale-dependent suppression of the correlation function at scales $r\lesssim 80 \Mpch$ to accommodate for the growing peak. We see a somewhat different behaviour for CDM in the middle panel of \reffig{CF-bc}: from $z=39$ to $z=7$, we see the BAO peak slightly increasing as CDM particles fall in the baryon potential well on these scales, imprinting the feature from the baryon field into the CDM field gradually (note that we observe the same position of the peak in baryons and CDM). The peak reaches its maximum relative amplitude at roughly $z=7$, the moment at which mild nonlinear effects appear. At redshift lower than $z=7$, we observe then a small decrease in the peak amplitude. On small scales, we note the same scale-dependent suppression for CDM fluctuations that appeared as well in the baryon fluctuations. The results here are in agreement with the ones in figure 9 of \cite{Angulo:2013qp}. In addition, as we saw for the halo-halo 2PCF in \reffig{void-halo-halo-particle-field} (green curves), and also for the halo-halo power spectrum in figure 9 of \cite{Khoraminezhad:2020zqe}, baryon-CDM relative perturbations tend to diminish the clustering. We however observed a slight increase of clustering on scales around the position of the BAO peak in the matter-matter 2PCF in \reffig{CF-real-zdepend}. We can now understand this in light of \reffig{CF-bc}: the pronounced baryon BAO feature increases the total matter BAO peak in 2-fluid simulations.

Finally, we compute the 2-point correlation function of the baryon-CDM perturbation field $\delta_{bc}$ in the bottom panel of \reffig{CF-bc}. We show this 2PCF only down to redshift $z=3$ because the noise increases as we reach lower redshift, and the 2PCF becomes consistent with zero on all scales. We see that this 2PCF is roughly constant close to zero, except for the BAO feature which is a BAO dip instead of the BAO peak in this case. This is because the BAO feature in the baryon field gradually imprints itself into the CDM field, which creates a skewed distribution of CDM with a sharp fall inside the BAO scale but with a larger tail on scales slightly larger than the BAO one (even though the position of the BAO peak is observed to be identical for baryons and CDM). Therefore we expect to observe an anti-correlation signal for $\delta_{bc}$ on scales slightly larger than the BAO scale ($\delta_{bc}$ is too small) in a skewed way, as can be seen in the bottom panel of \reffig{CF-bc}. We do not observe any notable redshift evolution for this 2PCF which is consistent with the fact that $\dbc$ itself is constant in time, as discussed in e.g. \cite{Schmidt:2016,Hahn:2020lvr,Khoraminezhad:2020zqe}. Notice that this kind of correlation function was also predicted using 2LPT in \cite{Chen:2019cfu}. While we do not conduct a detailed quantitative comparison of their prediction with our results, we note that they found the same kind of dip for correlation functions including relative baryon-CDM density perturbations.

\begin{figure*}
\centering
\includegraphics[scale=0.3]{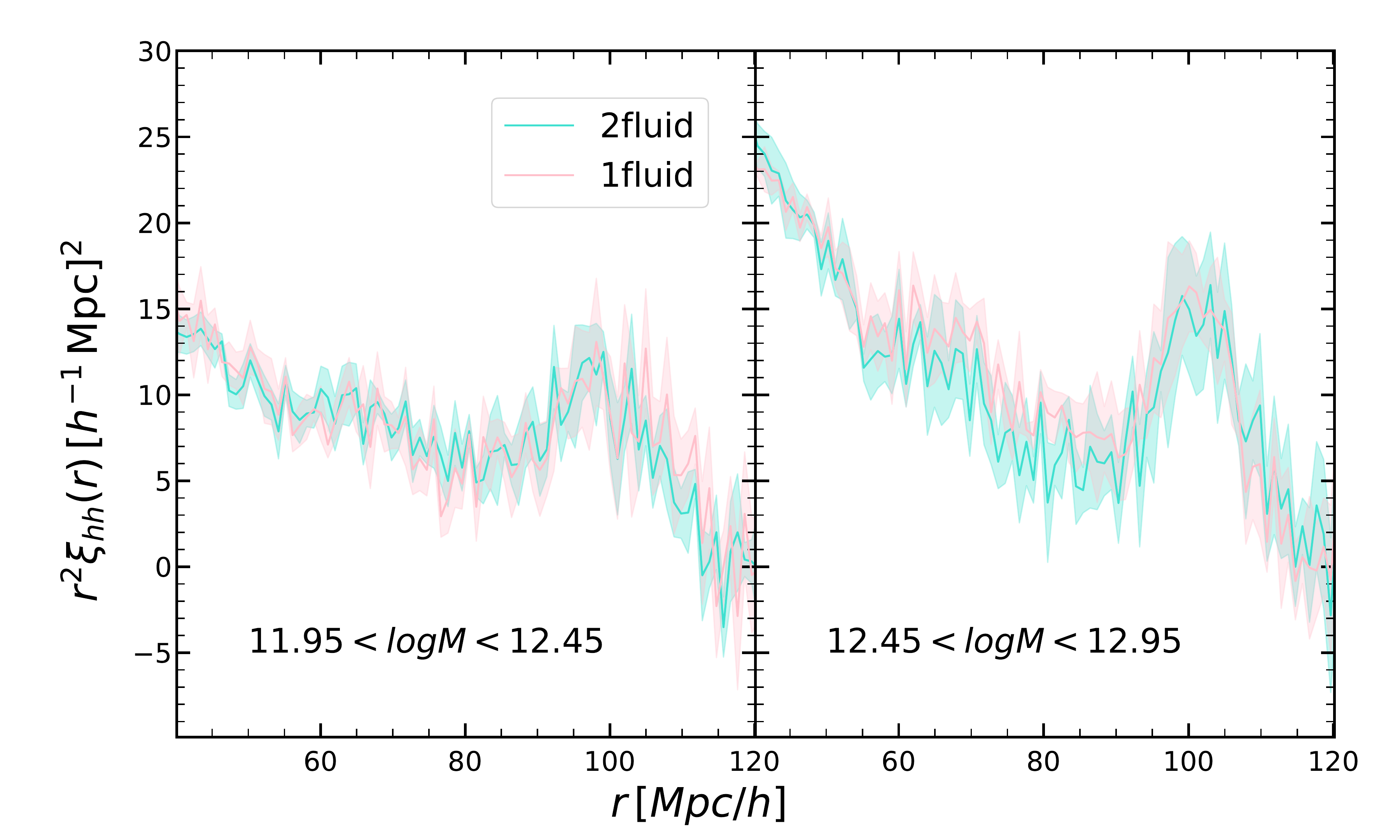}
\caption{Halo-halo 2-point correlation function at $z=0$ for 1-fluid (in pink) and 2-fluid (in blue) simulations for two different halo mass bins. The shaded area show the $1\sigma$ error over 8 realizations of each set of simulations. Again we multiply the 2PCF by $r^2$ to better see the BAO feature. We do not detect any impact of baryon-CDM relative perturbations on this 2PCF either.}
\label{fig:CF-hh}
\end{figure*}

Finally, we investigate the halo-halo 2PCF at redshift zero in \reffig{CF-hh}. We present results for two halo mass bins centered around $\log M = 12.2 M_\odot/h$ and $\log M = 12.7 M_\odot/h$. Recall that halos in the 2-fluid simulations are identified by considering both types of particles (baryons and CDM). As we see the halo-halo 2PCF is more noisy than the one obtained from particles due to the lower number of halos in comparison to particles. We see that results obtained in the 2-fluid simulations are fully consistent with the ones from 1-fluid simulations. This once again reflects the smallness of the impact of baryon-CDM perturbations on galaxy clustering at low redshift and implies that these effects will probably not need to be included in the modeling of correlation functions for the analysis of future surveys BAO peak estimation. This has a positive impact for such analysis since it will reduce the number of free parameters entering the model. These results are in line with previous results in the literature: \cite{Barreira:2019qdl, Khoraminezhad:2020zqe} estimated that the impact of baryon-CDM perturbations on the late-time halo power spectrum should not exceed 1 -- a few percent; \cite{Beutler:2016zat} conducted an analysis of the BOSS DR12 data with a model including baryon-CDM relative density and velocity perturbations, and obtained results for the bias parameters associated with such perturbations consistent with zero, indicating an effect too small to be detected; finally, using 2LPT, \cite{Chen:2019cfu} showed the effect to be at most one order of magnitude smaller than the halo 2PCF itself.

\subsection{Position of the BAO peak}
\label{sec:peak}

We now focus more specifically on the position of the BAO peak estimation for our two sets of simulation. \cite{Anselmi:2017zss} showed that the position of the BAO linear point, namely the midpoint scale between the peak and the dip of the 2PCF, can be extracted from the 2PCF measured in N-body simulations or galaxy data sets in a model-independent way by introducing a polynomial function to smooth the 2-point correlation function, and using a root-finding algorithm to estimate the zero-crossing of the first derivative of the 2PCF. To measure the linear point one needs to estimate the position of the BAO peak as well as the BAO dip through this modeling, but here we will just focus on the maximum of this fit. We use the following polynomial fit 
\be
\xi^{\rm fit} (r) = \sum_{n=0}^{N} a_{n}r^{n}.
\label{eq:poly}
\ee
Following \cite{Anselmi:2017zss}, we obtain the best fit parameter for the degree of the polynomial $N$ by minimizing the $\chi^2$. We use scales in the range $85 - 115 \Mpch$, and we choose $N=7$, which allows us to obtain good fits in the sense that the reduced $\chi^2$ is close to $1$ for all correlation functions we consider here while avoiding overfitting. We have also checked that the results for the position of the BAO peak depend only weakly on the degree of the polynomial (for example, the results for the matter-matter correlation function are consistent for polynomials of degree 4 to 8). Having the polynomial fit, to identify the peak position, we find the point the fit where the first derivative of the 2PCF is equal to zero, and the second derivative is negative.  

\begin{figure*}
\centering
\includegraphics[scale=0.3]{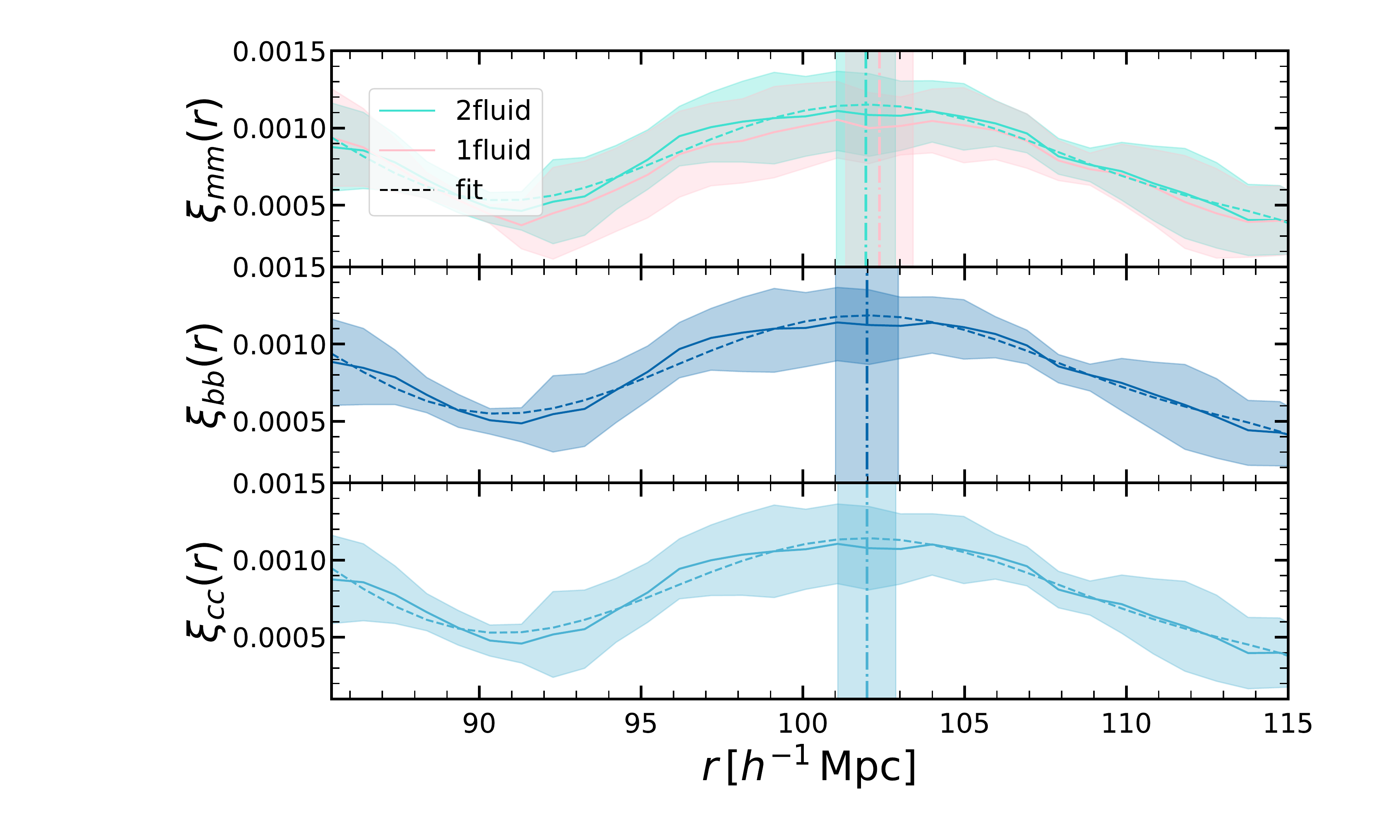}
\caption{\textbf{Top panel:} Comparison between the matter-matter 2PCF in 1-fluid (pink) and 2-fluid (cyan) simulations. \textbf{Middle and bottom panels:} baryon-baryon and CDM-CDM 2-point correlation functions respectively in 2-fluid simulations. Each time we show the 2PCF in real space at $z=0$, computed using \refeq{2pcf-sim}. The shaded area show the $1\sigma$ error over 8 realizations in each simulation. The vertical dotted-dashed lines show the position of the BAO peak obtained from a $7^{\rm{th}}$ degree polynomial fit of the form of \refeq{poly}. The polynomial fit is plotted in each case with the dashed line style and the same color for each type of correlations. The position of the BAO peak in each case for all different redshifts are presented in \reftab{baopeak}.}
\label{fig:CF-real-z0}
\end{figure*}

\refFig{CF-real-z0} illustrates this process by showing the matter-matter, CDM-CDM and baryon-baryon correlation functions, and the related position of the BAO peak in each case (dotted-dashed vertical line with $1\sigma$ error) at $z=0$. Each time the solid line shows the measurement while the dashed line shows the fit. We see that the position of the peak extracted from the baryon-baryon and CDM-CDM 2PCFs align with each other and with the total matter one in 2-fluid simulations. The position of the peak in 1-fluid simulation is slightly higher but the difference between the two cases lies within the 1$\sigma$ errorbars. As we explained in the discussion of \reffig{CF-bc}, this is expected since the BAO feature originates in the baryon field through baryon oscillations sourced by photon pressure, and then is imprinted into the CDM field with the same position but a lightly skewed distribution towards higher values. This results in a slightly overestimated position of the peak when assuming that the two fluids perfectly comove as is done in 1-fluid simulations.    

\begin{table*}
\centering
\begin{tabular}{|l|c|c|c|c|c|c|c|}
\hline
\hline
2fluid & redshift & CDM & baryon & $\delta_{bc}$& total matter & halo ($12.45 <\rm{log M} <12.95$)  \\ \hline\hline
       & $z=0$    &  $102.0\pm0.9$   &    $102.0\pm1.0$ & $102.3\pm5.7$  &     $102.0\pm0.9$& $101.2\pm3.5$     \\  
      \cline{2-3} \cline{3-4} \cline{4-5} \cline{5-6} \cline{6-7} \cline{7-8}  
       & $z=0.5$    &  $102.8\pm1.5$   &     $102.8\pm1.5$   &     $99.7\pm7.5$       &  $102.9\pm1.5$ &  $101.4\pm1.8$     \\
 \cline{2-3} \cline{3-4} \cline{4-5} \cline{5-6} \cline{6-7} \cline{7-8}
       & $z=1$    &   $102.9\pm1.2$  &    $103.0\pm1.1$    &     $105.2\pm2.8$      &  $102.9\pm1.2$  &  $104.5\pm4.9$ &        \\
 \cline{2-3} \cline{3-4} \cline{4-5} \cline{5-6} \cline{6-7} \cline{7-8}
       & $z=1.5$    &  $102.9\pm0.9$   &   $102.9\pm0.8$     &     $106.5\pm2.0$       &  $102.9\pm0.9$ &  $104.9\pm2.3$ &          \\
 \cline{2-3} \cline{3-4} \cline{4-5} \cline{5-6} \cline{6-7} \cline{7-8}
       & $z=2$    &  $102.7\pm0.7$   &     $102.8\pm0.7$   &      $104.7\pm1.5$     &  $102.7\pm0.7$  & $107.4\pm2.1$  &          \\ 
\cline{2-3} \cline{3-4} \cline{4-5} \cline{5-6} \cline{6-7} \cline{7-8}
       & $z=3$    &   $102.5\pm0.5$  &   $102.5\pm0.5$     &     $106.2\pm2.6$     &  $102.5\pm0.5$   & $105.9\pm2.8$  &         \\ 
       \hline\hline
1fluid & redshift & CDM & baryon &$\delta_{bc}$ & total matter & halo ($12.45 <\rm{log M} <12.95$) \\ \hline\hline
    & $z=0$    &  $-$   &   $-$     &  $-$   &   $102.4\pm1.0$       & $99.1\pm9.7$       \\
\cline{2-3} \cline{3-4} \cline{4-5} \cline{5-6} \cline{6-7} \cline{7-8}
       & $z=0.5$    &  $-$   &    $-$    & $-$  &      $103.0\pm1.8$      & $102.2\pm1.8$      \\
\cline{2-3} \cline{3-4} \cline{4-5} \cline{5-6} \cline{6-7} \cline{7-8}
       & $z=1$    &  $-$   &    $-$    &      $-$  &    $102.9\pm1.2$     & $101.9\pm3.5$  &          \\ \cline{2-3} \cline{3-4} \cline{4-5} \cline{5-6} \cline{6-7} \cline{7-8}
      & $z=1.5$    &  $-$   &    $-$    &      $-$  &    $102.8\pm0.8$     & $104.9\pm1.9$  &          \\
      \cline{2-3} \cline{3-4} \cline{4-5} \cline{5-6} \cline{6-7} \cline{7-8}
      & $z=2$    &   $-$  &    $-$    &       $-$  &    $102.7\pm0.7$    & $105.6\pm2.1$  &          \\
      \cline{2-3} \cline{3-4} \cline{4-5} \cline{5-6} \cline{6-7} \cline{7-8}
      & $z=3$    &  $-$   &    $-$    &       $-$  &    $102.5\pm0.5$    &  $106.4\pm1.9$ &          \\
      \hline\hline
\end{tabular}
\caption{Position of the BAO peak of the halo and matter fields in 1-fluid and 2-fluid simulations for different redshifts. In the case of 2-fluid set, we also compute the position of the peak for CDM, baryons, and the $\delta_{bc}$ fields separately. We see that any shift in the peak position is within $1\sigma$ errorbars.}
\label{table:baopeak}
\end{table*}

\begin{figure*}
\centering
\includegraphics[scale=0.3]{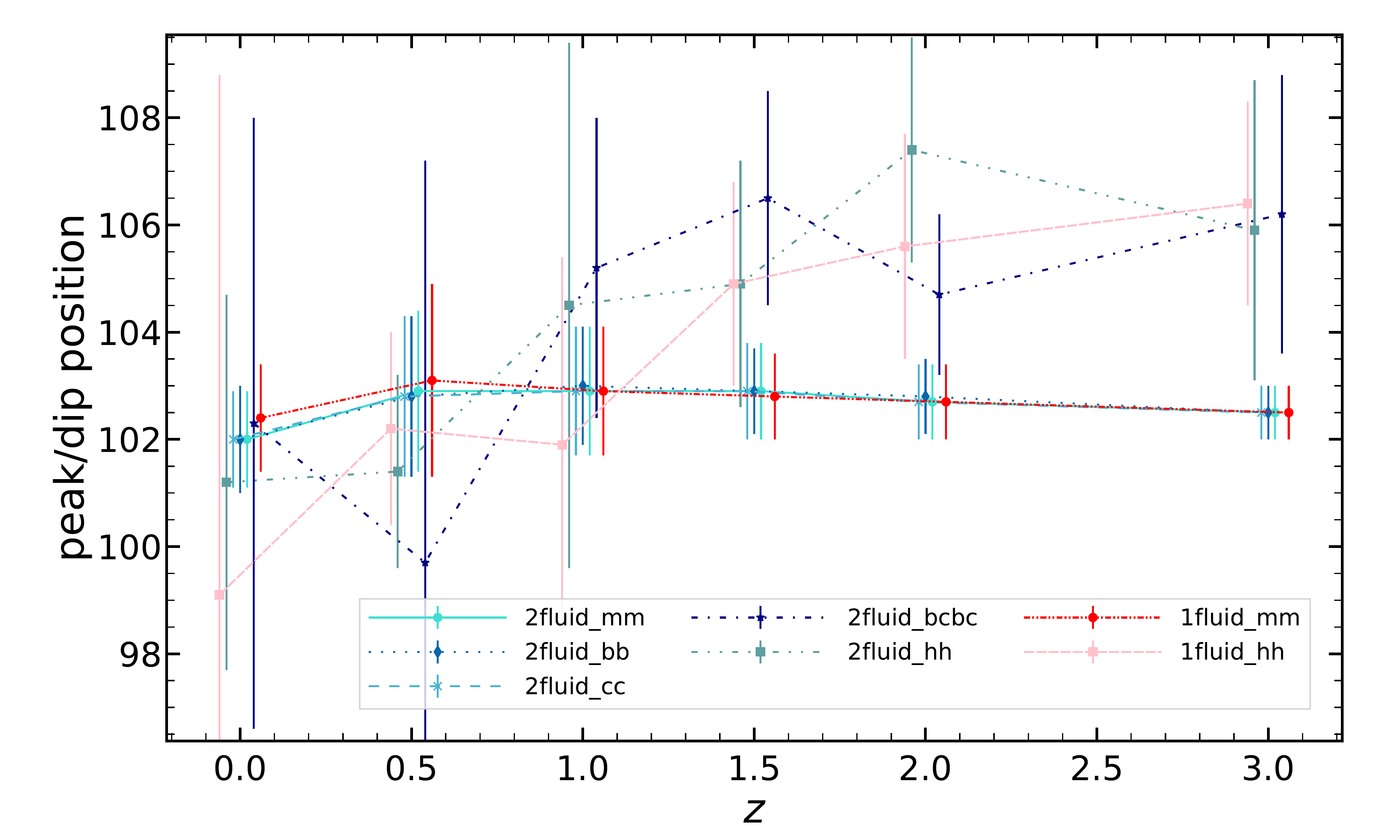}
\caption{Position of the BAO peak (or dip) for 1- and 2-fluid simulations for different fields for different redshifts. These results correspond to those of \reftab{baopeak}. 2-fluid measurements are shown in nuances of blue according to the legend, and 1-fluid ones in red. We present the matter-matter case with circle markers and the halo-halo case with square markers. We see no statistically significant differences between these two cases. The points have been slightly displaced horizontally to increase clarity. Each field is shown with a different line style.}
\label{fig:pos-z}
\end{figure*}

The values of the position of the BAO peak for each fluid and for several redshift are reported in \reftab{baopeak} as well as in \reffig{pos-z} in details. \refFig{pos-z} shows the 2-fluid measurements in nuances in blue for different fields and the 1-fluid case in red. We note that in the case of bcbc we show the position of the BAO dip both in \reffig{pos-z} and \reftab{baopeak}. Since the position of the BAO peak in all cases remains the same within errorbars (at least with the 8 realizations that we used here), we can argue that the BAO peak remains a standard ruler even in the presence of baryon-CDM perturbations. Notice that to decrease the errorbars by at least a factor of $5$, we would need at least $900$ realizations of each types of simulations but this would still not assure that we would see any significant differences. 

Using the results in \reffig{CF-hh}, we also computed the BAO peak position in the halo-halo 2-point correlation function for 2-fluid and 1-fluid simulations for the high mass bin. The results are shown in the last column of \reftab{baopeak}. As was already visible in the left panel of \reffig{CF-hh}, the positions of the peak are compatible within errorbars between the two cases.

Finally, we compute the position of the BAO feature in the bc-bc cross-correlation function from \reffig{CF-bc}. In this case we use the same polynomial fitting formula (\refeq{poly}) but looking now for the minimum of our fit. As we mentioned before, results at low $z$ become noisy which is why the errorbars on the peak position increase. The results are shown in the fourth column of \reftab{baopeak}. We do observe a somewhat higher value of the scale of the BAO dip with respect to that of the BAO peak of all other fields we consider (except halos), which is expected as explained before. 

To conclude, from \reftab{baopeak}, we do not detect any significant impact of relative baryon-CDM perturbations on the BAO peak position measured from the matter or halo correlation function. This is in line with results from the previous section where we found no evidence for a change in the broadband correlation function from such perturbations. This is also again in line with previous results from \cite{Beutler:2016zat} who found no evidence for nonzero bias parameters associated to these perturbations from the BOSS galaxy power spectrum. Furthermore, \cite{Barreira:2019qdl} also forecasted that the BAO peak position should be shifted by less than $1\%$ for halo samples similar to the one we consider here (their section 4). 

We end this section by a small word of caution. In this work we only considered the effects of baryon-CDM relative perturbations generated by baryon-photon coupling prior to recombination. However, as we already mentioned in the introduction, compensated isocurvature perturbations (CIP) can also be generated in some Inflation scenarios. As was discussed in \cite{Heinrich:2019sxl,Barreira:2020lva}, such CIPs can also locally affect the position of the BAO peak or the galaxy power spectrum, and these statistics could hence be used to constrain them as well as inflationary scenarios. A direct measurement of the impact of CIPs on the BAO peak position could be done using 1-fluid separate universe simulations as described in \cite{Barreira:2019qdl,Khoraminezhad:2020zqe}, but this is beyond the scope of this work.

\section{Summary and Conclusion}
\label{sec:conclude}

In this paper we performed 2-fluid gravity-only simulations building on our previous work in \cite{Khoraminezhad:2020zqe}, to study the impact of baryon-CDM relative perturbations due to photon pressure prior to recombination on voids statistics, density profile and clustering, as well as the 2PCF and position of the BAO peak in real space of various fluid components. The main findings of this study can be summarized as follows: 
\begin{itemize}
\item The VSF depends strongly on the tracer used to identify voids (there are more small voids and less large ones in the particle field than in the halo field). The VSF of particle field voids is unaffected by baryon-CDM relative perturbations, while the VSF of halo field voids is affected at $1 -2 \%$ level: smaller voids are more abundant in presence of such perturbations and larger voids less, which is a consequence of the fact that these perturbations act against clustering (\reffig{VSF_halo_particle}). 
\item We did not detect any statistically significant impact of baryon-CDM relative perturbations on the void, matter or halo auto- and cross- 2PCF. We found hints that these perturbations diminish the clustering on scales smaller than the BAO one, and enhance the BAO peak amplitude (\reffig{void-halo-halo-particle-field} and \reffigs{CF-real-zdepend}{CF-hh}), which is in agreement with our expectations. 
\item The density profiles of voids in halo and particle fields display the three known regimes (negative deep inside the void followed by the void profile regime with the positive compensation wall, and the linear regime where the halo-void correlation function becomes zero), and voids in the halo field are larger on average. We found no significant impact of baryon-CDM relative perturbations on any of the profiles, but a hint for voids in 2-fluid simulations to be emptier (\reffigs{void-profile_all}{void-profile_bins_particle}).  
\item The void bias depends significantly on the tracer used to find voids (the bias is almost constant over void size for halo field voids but it decreases for larger voids in the particle field), but we found consistent results for bias obtained from cross- and auto- correlation functions. Again we did not find any significant difference for the bias in 1- and 2-fluid simulations, but found hints that it is slightly larger in the latter case, as we expect (\reffig{void-bias-halo}).
\item The amplitude of the BAO peak in the baryon 2PCF decreases with time due to gravitational evolution. It is gradually imprinted in the CDM 2PCF where the amplitude of the peak grows down to $z\sim 7$ and then decreases down to $z=0$ due to nonlinear effects (\reffig{CF-bc}).
\item The relative density perturbation $\delta_{bc}$ auto-correlation function presents a dip as BAO feature on scales slightly larger than the BAO peak, which is consistent with the fact that on these scales CDM particles lag behind baryons (\reffig{CF-bc}).
\item We directly measured the impact that baryon-CDM perturbations have on the BAO peak position of halo and matter for the first time to our knowledge, and found no evidence for a statistically significant impact (\reffig{CF-real-z0}, \reffig{pos-z} and \reftab{baopeak}), which is in agreement with previous works (\cite{Beutler:2016zat}).
\end{itemize}

The halo field VSF is the only quantity that we found to be affected with statistical significance by baryon-CDM relative perturbations due to photon pressure prior to recombination. This effect might hence also affect the VSF of voids obtained from galaxy fields in observational data, and this statistics could hence be used to constraint such perturbations. We note however that the effect remains quite small. Our results for the matter-matter and halo-halo 2PCF added to ones from previous works confirm that the impact of baryon-CDM perturbations on cosmological constraints from the BAO feature in current and future galaxy surveys should be negligible at low redshift ($z \leq 3$). This has important consequences for future galaxy clustering surveys since it means that these effects will not have to be included in the modeling of leading-order quantities used for the analysis of their data. 

Finally, in the future, it would be interesting to use our extended set of simulations to reproduce the analysis in \cite{Khoraminezhad:2020zqe} including the two leading-order relative velocity bias parameters. This would allow to constraint their amplitude and their impact on the galaxy power spectrum. It would also be interested to reproduce the present study, at least partially,  using separate universe simulations described in \cite{Barreira:2019qdl} in order to measure the impact of CIPs generated during Inflation on voids statistics and BAO.

\section*{Acknowledgements}
We thank Gabriele Parimbelli, Tommaso Ronconi and Alexandre Barreira for useful discussions. HK, TL, MV, CB are supported by INFN INDARK grant. MV, CB also acknowledge contribution from the agreement ASI-INAF n.2017-14-H.0 and the ASI contracts Euclid-IC (I/031/10/0). CB, PV acknowledge support from the grant MIUR PRIN 2015 "Cosmology and Fundamental Physics: illuminating the Dark Universe with Euclid". CB also acknowledges support from the ASI COSMOS and LiteBIRD Networks (cosmosnet.it).

\section*{Data Availability Statement}
The data underlying this article will be shared on reasonable request to the corresponding author.

\bibliographystyle{mnras}
\bibliography{biblio} 

\begin{thebibliography}{}
\makeatletter
\relax
\def\mn@urlcharsother{\let\do\@makeother \do\$\do\&\do\#\do\^\do\_\do\%\do\~}
\def\mn@doi{\begingroup\mn@urlcharsother \@ifnextchar [ {\mn@doi@}
  {\mn@doi@[]}}
\def\mn@doi@[#1]#2{\def\@tempa{#1}\ifx\@tempa\@empty \href
  {http://dx.doi.org/#2} {doi:#2}\else \href {http://dx.doi.org/#2} {#1}\fi
  \endgroup}
\def\mn@eprint#1#2{\mn@eprint@#1:#2::\@nil}
\def\mn@eprint@arXiv#1{\href {http://arxiv.org/abs/#1} {{\tt arXiv:#1}}}
\def\mn@eprint@dblp#1{\href {http://dblp.uni-trier.de/rec/bibtex/#1.xml}
  {dblp:#1}}
\def\mn@eprint@#1:#2:#3:#4\@nil{\def\@tempa {#1}\def\@tempb {#2}\def\@tempc
  {#3}\ifx \@tempc \@empty \let \@tempc \@tempb \let \@tempb \@tempa \fi \ifx
  \@tempb \@empty \def\@tempb {arXiv}\fi \@ifundefined
  {mn@eprint@\@tempb}{\@tempb:\@tempc}{\expandafter \expandafter \csname
  mn@eprint@\@tempb\endcsname \expandafter{\@tempc}}}

\bibitem[\protect\citeauthoryear{Ade et~al.,}{Ade et~al.}{2016}]{Planck2015}
Ade P. A.~R.,  et~al., 2016, \mn@doi [Astronomy & Astrophysics]
  {10.1051/0004-6361/201525831}, 594, A21

\bibitem[\protect\citeauthoryear{Aghanim et~al.}{Aghanim
  et~al.}{2020}]{Aghanim:2018eyx}
Aghanim N.,  et~al., 2020, \mn@doi [Astron. Astrophys.]
  {10.1051/0004-6361/201833910}, 641, A6

\bibitem[\protect\citeauthoryear{Ahn}{Ahn}{2016}]{Ahn:2016bcr}
Ahn K.,  2016, \mn@doi [Astrophys. J.] {10.3847/0004-637X/830/2/68}, 830, 68

\bibitem[\protect\citeauthoryear{Alam et~al.}{Alam
  et~al.}{2021}]{eBOSS:2020yzd}
Alam S.,  et~al., 2021, \mn@doi [Phys. Rev. D] {10.1103/PhysRevD.103.083533},
  103, 083533

\bibitem[\protect\citeauthoryear{{Alcock} \& {Paczynski}}{{Alcock} \&
  {Paczynski}}{1979}]{Alcock_1979}
{Alcock} C.,  {Paczynski} B.,  1979, \mn@doi [Nature]
  {https://doi.org/10.1038/281358a0}, 281, {358}

\bibitem[\protect\citeauthoryear{Angulo \& Pontzen}{Angulo \&
  Pontzen}{2016}]{Angulo:2016hjd}
Angulo R.~E.,  Pontzen A.,  2016, \mn@doi [Mon. Not. Roy. Astron. Soc.]
  {10.1093/mnrasl/slw098}, 462, L1

\bibitem[\protect\citeauthoryear{Angulo, Hahn  \& Abel}{Angulo
  et~al.}{2013}]{Angulo:2013qp}
Angulo R.~E.,  Hahn O.,   Abel T.,  2013, \mn@doi [Mon. Not. Roy. Astron. Soc.]
  {10.1093/mnras/stt1135}, 434, 1756

\bibitem[\protect\citeauthoryear{Anselmi, Corasaniti, Starkman, Sheth  \&
  Zehavi}{Anselmi et~al.}{2018}]{Anselmi:2017zss}
Anselmi S.,  Corasaniti P.-S.,  Starkman G.~D.,  Sheth R.~K.,   Zehavi I.,
  2018, \mn@doi [Phys. Rev. D] {10.1103/PhysRevD.98.023527}, 98, 023527

\bibitem[\protect\citeauthoryear{{Baccigalupi}}{{Baccigalupi}}{1999}]{Baccigalupi_1999}
{Baccigalupi} C.,  1999, \mn@doi [\prd] {10.1103/PhysRevD.59.123004}, \href
  {https://ui.adsabs.harvard.edu/abs/1999PhRvD..59l3004B} {59, 123004}

\bibitem[\protect\citeauthoryear{Baccigalupi, Amendola  \&
  Occhionero}{Baccigalupi et~al.}{1997}]{Baccigalupi_1997}
Baccigalupi C.,  Amendola L.,   Occhionero F.,  1997, \mn@doi [Monthly Notices
  of the Royal Astronomical Society] {10.1093/mnras/288.2.387}, 288, 387

\bibitem[\protect\citeauthoryear{Baldi \& Villaescusa-Navarro}{Baldi \&
  Villaescusa-Navarro}{2018}]{Baldi:2016oce}
Baldi M.,  Villaescusa-Navarro F.,  2018, \mn@doi [Mon. Not. Roy. Astron. Soc.]
  {10.1093/mnras/stx2594}, 473, 3226

\bibitem[\protect\citeauthoryear{Barkana \& Loeb}{Barkana \&
  Loeb}{2011}]{Barkana:2010zq}
Barkana R.,  Loeb A.,  2011, \mn@doi [Mon. Not. Roy. Astron. Soc.]
  {10.1111/j.1365-2966.2011.18922.x}, 415, 3113

\bibitem[\protect\citeauthoryear{Barreira, Cautun, Li, Baugh  \&
  Pascoli}{Barreira et~al.}{2015}]{Barreira:2015vra}
Barreira A.,  Cautun M.,  Li B.,  Baugh C.,   Pascoli S.,  2015, \mn@doi [JCAP]
  {10.1088/1475-7516/2015/08/028}, 08, 028

\bibitem[\protect\citeauthoryear{Barreira, Cabass, Nelson  \& Schmidt}{Barreira
  et~al.}{2020a}]{Barreira:2019qdl}
Barreira A.,  Cabass G.,  Nelson D.,   Schmidt F.,  2020a, \mn@doi [JCAP]
  {10.1088/1475-7516/2020/02/005}, 02, 005

\bibitem[\protect\citeauthoryear{Barreira, Cabass, Lozanov  \&
  Schmidt}{Barreira et~al.}{2020b}]{Barreira:2020lva}
Barreira A.,  Cabass G.,  Lozanov K.~D.,   Schmidt F.,  2020b, \mn@doi [JCAP]
  {10.1088/1475-7516/2020/07/049}, 07, 049

\bibitem[\protect\citeauthoryear{Beutler, Seljak  \& Vlah}{Beutler
  et~al.}{2017}]{Beutler:2016zat}
Beutler F.,  Seljak U.,   Vlah Z.,  2017, \mn@doi [Mon. Not. Roy. Astron. Soc.]
  {10.1093/mnras/stx1196}, 470, 2723

\bibitem[\protect\citeauthoryear{Bird, Feng, Pedersen  \& Font-Ribera}{Bird
  et~al.}{2020}]{Bird:2020}
Bird S.,  Feng Y.,  Pedersen C.,   Font-Ribera A.,  2020, \mn@doi [JCAP]
  {10.1088/1475-7516/2020/06/002}, 06, 002

\bibitem[\protect\citeauthoryear{Blazek, McEwen  \& Hirata}{Blazek
  et~al.}{2016}]{Blazek:2015ula}
Blazek J.,  McEwen J.~E.,   Hirata C.~M.,  2016, \mn@doi [Phys. Rev. Lett.]
  {10.1103/PhysRevLett.116.121303}, 116, 121303

\bibitem[\protect\citeauthoryear{Bos, van~de Weygaert, Dolag  \& Pettorino}{Bos
  et~al.}{2012}]{Bos:2012wq}
Bos E. G.~P.,  van~de Weygaert R.,  Dolag K.,   Pettorino V.,  2012, \mn@doi
  [Mon. Not. Roy. Astron. Soc.] {10.1111/j.1365-2966.2012.21478.x}, 426, 440

\bibitem[\protect\citeauthoryear{Cai, Neyrinck, Szapudi, Cole  \& Frenk}{Cai
  et~al.}{2014}]{Cai2014}
Cai Y.-C.,  Neyrinck M.~C.,  Szapudi I.,  Cole S.,   Frenk C.~S.,  2014,
  \mn@doi [The Astrophysical Journal] {10.1088/0004-637x/786/2/110}, 786, 110

\bibitem[\protect\citeauthoryear{Cai, Padilla  \& Li}{Cai
  et~al.}{2015}]{Cai:2014fma}
Cai Y.-C.,  Padilla N.,   Li B.,  2015, \mn@doi [Mon. Not. Roy. Astron. Soc.]
  {10.1093/mnras/stv777}, 451, 1036

\bibitem[\protect\citeauthoryear{{Cai}, {Neyrinck}, {Mao}, {Peacock}, {Szapudi}
   \& {Berlind}}{{Cai} et~al.}{2017}]{cai2017}
{Cai} Y.-C.,  {Neyrinck} M.,  {Mao} Q.,  {Peacock} J.~A.,  {Szapudi} I.,
  {Berlind} A.~A.,  2017, \mn@doi [\mnras] {10.1093/mnras/stw3299}, \href
  {https://ui.adsabs.harvard.edu/abs/2017MNRAS.466.3364C} {466, 3364}

\bibitem[\protect\citeauthoryear{{Cautun}, {van de Weygaert}, {Jones}  \&
  {Frenk}}{{Cautun} et~al.}{2014}]{10.1093/mnras/stu768}
{Cautun} M.,  {van de Weygaert} R.,  {Jones} B. J.~T.,   {Frenk} C.~S.,  2014,
  \mn@doi [\mnras] {10.1093/mnras/stu768}, \href
  {https://ui.adsabs.harvard.edu/abs/2014MNRAS.441.2923C} {441, 2923}

\bibitem[\protect\citeauthoryear{Ceccarelli, Paz, Lares, Padilla  \&
  Lambas}{Ceccarelli et~al.}{2013}]{Ceccarelli:2013rza}
Ceccarelli L.,  Paz D.,  Lares M.,  Padilla N.,   Lambas D.~G.,  2013, \mn@doi
  [Mon. Not. Roy. Astron. Soc.] {10.1093/mnras/stt1097}, 434, 1435

\bibitem[\protect\citeauthoryear{Chan, Hamaus  \& Desjacques}{Chan
  et~al.}{2014}]{Chan:2014qka}
Chan K.~C.,  Hamaus N.,   Desjacques V.,  2014, \mn@doi [Phys. Rev. D]
  {10.1103/PhysRevD.90.103521}, 90, 103521

\bibitem[\protect\citeauthoryear{{Chan}, {Hamaus}  \& {Biagetti}}{{Chan}
  et~al.}{2019}]{chan2019}
{Chan} K.~C.,  {Hamaus} N.,   {Biagetti} M.,  2019, \mn@doi [\prd]
  {10.1103/PhysRevD.99.121304}, \href
  {https://ui.adsabs.harvard.edu/abs/2019PhRvD..99l1304C} {99, 121304}

\bibitem[\protect\citeauthoryear{{Chan}, {Li}, {Biagetti}  \& {Hamaus}}{{Chan}
  et~al.}{2020}]{chan2020}
{Chan} K.~C.,  {Li} Y.,  {Biagetti} M.,   {Hamaus} N.,  2020, \mn@doi [\apj]
  {10.3847/1538-4357/ab64ec}, \href
  {https://ui.adsabs.harvard.edu/abs/2020ApJ...889...89C} {889, 89}

\bibitem[\protect\citeauthoryear{Chen, Castorina  \& White}{Chen
  et~al.}{2019}]{Chen:2019cfu}
Chen S.-F.,  Castorina E.,   White M.,  2019, \mn@doi [JCAP]
  {10.1088/1475-7516/2019/06/006}, 06, 006

\bibitem[\protect\citeauthoryear{Christopherson}{Christopherson}{2014}]{Christopherson:2014eoa}
Christopherson A.~J.,  2014, \mn@doi [Int. J. Mod. Phys. D]
  {10.1142/S0218271814300249}, 23, 1430024

\bibitem[\protect\citeauthoryear{Clampitt, Cai  \& Li}{Clampitt
  et~al.}{2013}]{10.1093/mnras/stt219}
Clampitt J.,  Cai Y.-C.,   Li B.,  2013, \mn@doi [Monthly Notices of the Royal
  Astronomical Society] {10.1093/mnras/stt219}, 431, 749

\bibitem[\protect\citeauthoryear{Clampitt, Jain  \& S\'anchez}{Clampitt
  et~al.}{2016}]{Clampitt:2015jra}
Clampitt J.,  Jain B.,   S\'anchez C.,  2016, \mn@doi [Mon. Not. Roy. Astron.
  Soc.] {10.1093/mnras/stv2933}, 456, 4425

\bibitem[\protect\citeauthoryear{Colberg, Sheth, Diaferio, Gao  \&
  Yoshida}{Colberg et~al.}{2005}]{Colberg:2004nd}
Colberg J.~M.,  Sheth R.~K.,  Diaferio A.,  Gao L.,   Yoshida N.,  2005,
  \mn@doi [Mon. Not. Roy. Astron. Soc.] {10.1111/j.1365-2966.2005.09064.x},
  360, 216

\bibitem[\protect\citeauthoryear{{Contarini}, {Ronconi}, {Marulli},
  {Moscardini}, {Veropalumbo}  \& {Baldi}}{{Contarini}
  et~al.}{2019}]{Contarini2019}
{Contarini} S.,  {Ronconi} T.,  {Marulli} F.,  {Moscardini} L.,  {Veropalumbo}
  A.,   {Baldi} M.,  2019, \mn@doi [\mnras] {10.1093/mnras/stz1989}, \href
  {https://ui.adsabs.harvard.edu/abs/2019MNRAS.488.3526C} {488, 3526}

\bibitem[\protect\citeauthoryear{Contarini, Marulli, Moscardini, Veropalumbo,
  Giocoli  \& Baldi}{Contarini et~al.}{2021}]{Contarini_2021}
Contarini S.,  Marulli F.,  Moscardini L.,  Veropalumbo A.,  Giocoli C.,
  Baldi M.,  2021, \mn@doi [Monthly Notices of the Royal Astronomical Society]
  {10.1093/mnras/stab1112}, 504, 5021–5038

\bibitem[\protect\citeauthoryear{D'Amico, Musso, Nore\~na  \&
  Paranjape}{D'Amico et~al.}{2011}]{PhysRevD.83.023521}
D'Amico G.,  Musso M.,  Nore\~na J.,   Paranjape A.,  2011, \mn@doi [Phys. Rev.
  D] {10.1103/PhysRevD.83.023521}, 83, 023521

\bibitem[\protect\citeauthoryear{Dalal, Pen  \& Seljak}{Dalal
  et~al.}{2010}]{Dalal_2010}
Dalal N.,  Pen U.-L.,   Seljak U.,  2010, \mn@doi [Journal of Cosmology and
  Astroparticle Physics] {10.1088/1475-7516/2010/11/007}, 2010, 007

\bibitem[\protect\citeauthoryear{Davis \& Peebles}{Davis \&
  Peebles}{1983}]{Davis_peebles_1983}
Davis M.,  Peebles P. J.~E.,  1983, \mn@doi [Astrophysical Journal]
  {10.1086/160884}, 267, 465

\bibitem[\protect\citeauthoryear{Desjacques, Jeong  \& Schmidt}{Desjacques
  et~al.}{2018}]{Desjacques:2016bnm}
Desjacques V.,  Jeong D.,   Schmidt F.,  2018, \mn@doi [Phys. Rept.]
  {10.1016/j.physrep.2017.12.002}, 733, 1

\bibitem[\protect\citeauthoryear{{Falck} \& {Neyrinck}}{{Falck} \&
  {Neyrinck}}{2015}]{10.1093/mnras/stv879}
{Falck} B.,  {Neyrinck} M.~C.,  2015, \mn@doi [\mnras] {10.1093/mnras/stv879},
  \href {https://ui.adsabs.harvard.edu/abs/2015MNRAS.450.3239F} {450, 3239}

\bibitem[\protect\citeauthoryear{Ferrer, Rasanen  \& Valiviita}{Ferrer
  et~al.}{2004}]{Ferrer:2004nv}
Ferrer F.,  Rasanen S.,   Valiviita J.,  2004, \mn@doi [JCAP]
  {10.1088/1475-7516/2004/10/010}, 10, 010

\bibitem[\protect\citeauthoryear{Finelli, García-Bellido, Kovács, Paci  \&
  Szapudi}{Finelli et~al.}{2014}]{finelli_2014}
Finelli F.,  García-Bellido J.,  Kovács A.,  Paci F.,   Szapudi I.,  2014,
  \mn@doi [Proceedings of the International Astronomical Union]
  {https://doi.org/10.1017/S1743921314013714}, 10, 153–155

\bibitem[\protect\citeauthoryear{{Furlanetto} \& {Piran}}{{Furlanetto} \&
  {Piran}}{2006}]{Furlanetto2006}
{Furlanetto} S.~R.,  {Piran} T.,  2006, \mn@doi [\mnras]
  {10.1111/j.1365-2966.2005.09862.x}, \href
  {https://ui.adsabs.harvard.edu/abs/2006MNRAS.366..467F} {366, 467}

\bibitem[\protect\citeauthoryear{Gibbons, Werner, Yoshida  \& Chon}{Gibbons
  et~al.}{2013}]{10.1093/mnras/stt2298}
Gibbons G.~W.,  Werner M.~C.,  Yoshida N.,   Chon S.,  2013, \mn@doi [Monthly
  Notices of the Royal Astronomical Society] {10.1093/mnras/stt2298}, 438, 1603

\bibitem[\protect\citeauthoryear{Gill, Knebe  \& Gibson}{Gill
  et~al.}{2004}]{Gill:2004}
Gill S.~P.,  Knebe A.,   Gibson B.~K.,  2004, \mn@doi [Mon.Not.Roy.Astron.Soc.]
  {10.1111/j.1365-2966.2004.07786.x}, 351, 399

\bibitem[\protect\citeauthoryear{{Granett}, {Neyrinck}  \& {Szapudi}}{{Granett}
  et~al.}{2008}]{granett2008}
{Granett} B.~R.,  {Neyrinck} M.~C.,   {Szapudi} I.,  2008, \mn@doi [\apjl]
  {10.1086/591670}, \href
  {https://ui.adsabs.harvard.edu/abs/2008ApJ...683L..99G} {683, L99}

\bibitem[\protect\citeauthoryear{{Granett}, {Kov{\'a}cs}  \&
  {Hawken}}{{Granett} et~al.}{2015}]{granettt2015}
{Granett} B.~R.,  {Kov{\'a}cs} A.,   {Hawken} A.~J.,  2015, \mn@doi [\mnras]
  {10.1093/mnras/stv2110}, \href
  {https://ui.adsabs.harvard.edu/abs/2015MNRAS.454.2804G} {454, 2804}

\bibitem[\protect\citeauthoryear{Gregory, Thompson  \& Tifft}{Gregory
  et~al.}{1978}]{Gregory_1978}
Gregory S.~A.,  Thompson L.~A.,   Tifft W.~G.,  1978, BAAS, 10, 622

\bibitem[\protect\citeauthoryear{Grin, Dore  \& Kamionkowski}{Grin
  et~al.}{2011}]{Grin:2011}
Grin D.,  Dore O.,   Kamionkowski M.,  2011, \mn@doi [Phys. Rev. D]
  {10.1103/PhysRevD.84.123003}, 84, 123003

\bibitem[\protect\citeauthoryear{Hahn \& Abel}{Hahn \& Abel}{2011}]{Hahn_2011}
Hahn O.,  Abel T.,  2011, \mn@doi [Monthly Notices of the Royal Astronomical
  Society] {10.1111/j.1365-2966.2011.18820.x}, 415, 2101–2121

\bibitem[\protect\citeauthoryear{Hahn, Rampf  \& Uhlemann}{Hahn
  et~al.}{2021}]{Hahn:2020lvr}
Hahn O.,  Rampf C.,   Uhlemann C.,  2021, \mn@doi [Mon. Not. Roy. Astron. Soc.]
  {10.1093/mnras/staa3773}, 503, 426

\bibitem[\protect\citeauthoryear{Hamaus, Sutter  \& Wandelt}{Hamaus
  et~al.}{2014a}]{PhysRevLett.112.251302}
Hamaus N.,  Sutter P.~M.,   Wandelt B.~D.,  2014a, \mn@doi [Phys. Rev. Lett.]
  {10.1103/PhysRevLett.112.251302}, 112, 251302

\bibitem[\protect\citeauthoryear{Hamaus, Wandelt, Sutter, Lavaux  \&
  Warren}{Hamaus et~al.}{2014b}]{Hamaus:2013qja}
Hamaus N.,  Wandelt B.~D.,  Sutter P.~M.,  Lavaux G.,   Warren M.~S.,  2014b,
  \mn@doi [Phys. Rev. Lett.] {10.1103/PhysRevLett.112.041304}, 112, 041304

\bibitem[\protect\citeauthoryear{Hamaus, Sutter, Lavaux  \& Wandelt}{Hamaus
  et~al.}{2015}]{Hamaus:2015yza}
Hamaus N.,  Sutter P.~M.,  Lavaux G.,   Wandelt B.~D.,  2015, \mn@doi [JCAP]
  {10.1088/1475-7516/2015/11/036}, 11, 036

\bibitem[\protect\citeauthoryear{Hamaus, Pisani, Sutter, Lavaux, Escoffier,
  Wandelt  \& Weller}{Hamaus et~al.}{2016}]{Hamaus:2016wka}
Hamaus N.,  Pisani A.,  Sutter P.~M.,  Lavaux G.,  Escoffier S.,  Wandelt
  B.~D.,   Weller J.,  2016, \mn@doi [Phys. Rev. Lett.]
  {10.1103/PhysRevLett.117.091302}, 117, 091302

\bibitem[\protect\citeauthoryear{Hamaus et~al.,}{Hamaus
  et~al.}{2021}]{hamaus2021euclid}
Hamaus N.,  et~al., 2021, Euclid: Forecasts from redshift-space distortions and
  the Alcock-Paczynski test with cosmic voids (\mn@eprint {arXiv} {2108.10347})

\bibitem[\protect\citeauthoryear{Hamilton}{Hamilton}{1993}]{Hamilton_1993}
Hamilton A. J.~S.,  1993, \mn@doi [Astrophysical Journal] {10.1086/173288},
  417, 19

\bibitem[\protect\citeauthoryear{Hand, Feng, Beutler, Li, Modi, Seljak  \&
  Slepian}{Hand et~al.}{2018}]{Hand2018}
Hand N.,  Feng Y.,  Beutler F.,  Li Y.,  Modi C.,  Seljak U.,   Slepian Z.,
  2018, \mn@doi [The Astronomical Journal] {10.3847/1538-3881/aadae0}, 156, 160

\bibitem[\protect\citeauthoryear{{Hang}, {Alam}, {Cai}  \& {Peacock}}{{Hang}
  et~al.}{2021}]{hang2021}
{Hang} Q.,  {Alam} S.,  {Cai} Y.-C.,   {Peacock} J.,  2021, arXiv e-prints,
  \href {https://ui.adsabs.harvard.edu/abs/2021arXiv210511936H} {p.
  arXiv:2105.11936}

\bibitem[\protect\citeauthoryear{Hausman, Olson  \& Roth}{Hausman
  et~al.}{1983}]{Hausman_1983}
Hausman M.~A.,  Olson D.~W.,   Roth B.~D.,  1983, \mn@doi [The Astrophysical
  Journal] {https://doi.org/10.1086/161128}, 270, 351

\bibitem[\protect\citeauthoryear{He, Grin  \& Hu}{He et~al.}{2015}]{He:2015msa}
He C.,  Grin D.,   Hu W.,  2015, \mn@doi [Phys. Rev. D]
  {10.1103/PhysRevD.92.063018}, 92, 063018

\bibitem[\protect\citeauthoryear{Heinrich \& Schmittfull}{Heinrich \&
  Schmittfull}{2019}]{Heinrich:2019sxl}
Heinrich C.,  Schmittfull M.,  2019, \mn@doi [Phys. Rev. D]
  {10.1103/PhysRevD.100.063503}, 100, 063503

\bibitem[\protect\citeauthoryear{Hewett}{Hewett}{1982}]{Hewett_1982}
Hewett P.~C.,  1982, \mn@doi [Astrophysical Journal] {10.1093/mnras/201.4.867},
  201, 867

\bibitem[\protect\citeauthoryear{{Hotchkiss}, {Nadathur}, {Gottl{\"o}ber},
  {Iliev}, {Knebe}, {Watson}  \& {Yepes}}{{Hotchkiss}
  et~al.}{2015}]{Hotchkiss2015}
{Hotchkiss} S.,  {Nadathur} S.,  {Gottl{\"o}ber} S.,  {Iliev} I.~T.,  {Knebe}
  A.,  {Watson} W.~A.,   {Yepes} G.,  2015, \mn@doi [\mnras]
  {10.1093/mnras/stu2072}, \href
  {https://ui.adsabs.harvard.edu/abs/2015MNRAS.446.1321H} {446, 1321}

\bibitem[\protect\citeauthoryear{Hotinli, Mertens, Johnson  \&
  Kamionkowski}{Hotinli et~al.}{2019}]{Hotinli:2019wdp}
Hotinli S.~C.,  Mertens J.~B.,  Johnson M.~C.,   Kamionkowski M.,  2019,
  \mn@doi [Phys. Rev. D] {10.1103/PhysRevD.100.103528}, 100, 103528

\bibitem[\protect\citeauthoryear{Huston \& Christopherson}{Huston \&
  Christopherson}{2014}]{Huston:2013kgl}
Huston I.,  Christopherson A.~J.,  2014, Isocurvature Perturbations and
  Reheating in Multi-Field Inflation (\mn@eprint {arXiv} {1302.4298})

\bibitem[\protect\citeauthoryear{Iannuzzi \& Dolag}{Iannuzzi \&
  Dolag}{2011}]{Iannuzzi2011AdaptiveGS}
Iannuzzi F.,  Dolag K.,  2011, Monthly Notices of the Royal Astronomical
  Society, 417, 2846

\bibitem[\protect\citeauthoryear{Jamieson \& Loverde}{Jamieson \&
  Loverde}{2019a}]{Jamieson:2018biz}
Jamieson D.,  Loverde M.,  2019a, \mn@doi [Physical Review D]
  {10.1103/physrevd.100.023516}, 100

\bibitem[\protect\citeauthoryear{Jamieson \& Loverde}{Jamieson \&
  Loverde}{2019b}]{Jamieson:2019dmp}
Jamieson D.,  Loverde M.,  2019b, \mn@doi [Phys. Rev. D]
  {10.1103/PhysRevD.100.123528}, 100, 123528

\bibitem[\protect\citeauthoryear{Jeffrey et~al.,}{Jeffrey
  et~al.}{2021}]{DES:2021gua}
Jeffrey N.,  et~al., 2021, \mn@doi [Monthly Notices of the Royal Astronomical
  Society] {10.1093/mnras/stab1495}, 505, 4626–4645

\bibitem[\protect\citeauthoryear{Khoraminezhad, Lazeyras, Angulo, Hahn  \&
  Viela}{Khoraminezhad et~al.}{2021}]{Khoraminezhad:2020zqe}
Khoraminezhad H.,  Lazeyras T.,  Angulo R.~E.,  Hahn O.,   Viela M.,  2021,
  \mn@doi [JCAP] {10.1088/1475-7516/2021/03/023}, 03, 023

\bibitem[\protect\citeauthoryear{Kirshner, Oemler, Schechter  \&
  Shectman}{Kirshner et~al.}{1981}]{krishner_1981}
Kirshner R.~P.,  Oemler A. J.,  Schechter P.~L.,   Shectman S.~A.,  1981,
  \mn@doi [The Astrophysical Journal] {https://doi.org/10.1086/183623}, 248,
  L57

\bibitem[\protect\citeauthoryear{Knollmann \& Knebe}{Knollmann \&
  Knebe}{2009}]{Knollmann:2009}
Knollmann S.~R.,  Knebe A.,  2009, \mn@doi [Astrophys.J.Suppl.]
  {10.1088/0067-0049/182/2/608}, 182, 608

\bibitem[\protect\citeauthoryear{{Kov{\'a}cs} et~al.,}{{Kov{\'a}cs}
  et~al.}{2017}]{kovacs2017}
{Kov{\'a}cs} A.,  et~al., 2017, \mn@doi [\mnras] {10.1093/mnras/stw2968}, \href
  {https://ui.adsabs.harvard.edu/abs/2017MNRAS.465.4166K} {465, 4166}

\bibitem[\protect\citeauthoryear{{Kov{\'a}cs} et~al.,}{{Kov{\'a}cs}
  et~al.}{2019}]{kovacs2019}
{Kov{\'a}cs} A.,  et~al., 2019, \mn@doi [\mnras] {10.1093/mnras/stz341}, \href
  {https://ui.adsabs.harvard.edu/abs/2019MNRAS.484.5267K} {484, 5267}

\bibitem[\protect\citeauthoryear{Kovač et~al.,}{Kovač
  et~al.}{2013}]{10.1093/mnras/stt2241}
Kovač K.,  et~al., 2013, \mn@doi [Monthly Notices of the Royal Astronomical
  Society] {10.1093/mnras/stt2241}, 438, 717

\bibitem[\protect\citeauthoryear{Kreisch, Pisani, Carbone, Liu, Hawken,
  Massara, Spergel  \& Wandelt}{Kreisch et~al.}{2019}]{10.1093/mnras/stz1944}
Kreisch C.~D.,  Pisani A.,  Carbone C.,  Liu J.,  Hawken A.~J.,  Massara E.,
  Spergel D.~N.,   Wandelt B.~D.,  2019, \mn@doi [Monthly Notices of the Royal
  Astronomical Society] {10.1093/mnras/stz1944}, 488, 4413

\bibitem[\protect\citeauthoryear{{Landy} \& {Szalay}}{{Landy} \&
  {Szalay}}{1993}]{Landy1994}
{Landy} S.~D.,  {Szalay} A.~S.,  1993, \mn@doi [\apj] {10.1086/172900}, \href
  {https://ui.adsabs.harvard.edu/abs/1993ApJ...412...64L} {412, 64}

\bibitem[\protect\citeauthoryear{Langlois \& Riazuelo}{Langlois \&
  Riazuelo}{2000}]{Langlois:2000ar}
Langlois D.,  Riazuelo A.,  2000, \mn@doi [Phys. Rev. D]
  {10.1103/PhysRevD.62.043504}, 62, 043504

\bibitem[\protect\citeauthoryear{Lavaux \& Wandelt}{Lavaux \&
  Wandelt}{2010}]{Lavaux_2010}
Lavaux G.,  Wandelt B.~D.,  2010, \mn@doi [Monthly Notices of the Royal
  Astronomical Society] {10.1111/j.1365-2966.2010.16197.x}, 403, 1392–1408

\bibitem[\protect\citeauthoryear{Lavaux \& Wandelt}{Lavaux \&
  Wandelt}{2012}]{Lavaux_2012}
Lavaux G.,  Wandelt B.~D.,  2012, \mn@doi [The Astrophysical Journal]
  {10.1088/0004-637x/754/2/109}, 754, 109

\bibitem[\protect\citeauthoryear{Lewis, Challinor  \& Lasenby}{Lewis
  et~al.}{2000}]{Lewis_2000}
Lewis A.,  Challinor A.,   Lasenby A.,  2000, \mn@doi [The Astrophysical
  Journal] {https://doi.org/10.1086/309179}, 538, 473

\bibitem[\protect\citeauthoryear{Li}{Li}{2011}]{10.1111/j.1365-2966.2010.17867.x}
Li B.,  2011, \mn@doi [Monthly Notices of the Royal Astronomical Society]
  {10.1111/j.1365-2966.2010.17867.x}, 411, 2615

\bibitem[\protect\citeauthoryear{Li, Lin, Wang  \& Wang}{Li
  et~al.}{2009}]{Li:2008jn}
Li M.,  Lin C.,  Wang T.,   Wang Y.,  2009, \mn@doi [Phys. Rev. D]
  {10.1103/PhysRevD.79.063526}, 79, 063526

\bibitem[\protect\citeauthoryear{Liddle \& Mazumdar}{Liddle \&
  Mazumdar}{2000}]{Liddle:1999pr}
Liddle A.~R.,  Mazumdar A.,  2000, \mn@doi [Phys. Rev. D]
  {10.1103/PhysRevD.61.123507}, 61, 123507

\bibitem[\protect\citeauthoryear{Linde \& Mukhanov}{Linde \&
  Mukhanov}{1997}]{Linde:1996gt}
Linde A.~D.,  Mukhanov V.~F.,  1997, \mn@doi [Phys. Rev. D]
  {10.1103/PhysRevD.56.R535}, 56, 535

\bibitem[\protect\citeauthoryear{Lyth, Ungarelli  \& Wands}{Lyth
  et~al.}{2003}]{Lyth:2002my}
Lyth D.~H.,  Ungarelli C.,   Wands D.,  2003, \mn@doi [Phys. Rev. D]
  {10.1103/PhysRevD.67.023503}, 67, 023503

\bibitem[\protect\citeauthoryear{Mao et~al.}{Mao et~al.}{2017}]{Mao:2016onb}
Mao Q.,  et~al., 2017, \mn@doi [Astrophys. J.] {10.3847/1538-4357/835/2/161},
  835, 161

\bibitem[\protect\citeauthoryear{{Massara}, {Villaescusa-Navarro}, {Viel}  \&
  {Sutter}}{{Massara} et~al.}{2015}]{Massara:2015}
{Massara} E.,  {Villaescusa-Navarro} F.,  {Viel} M.,   {Sutter} P.~M.,  2015,
  \mn@doi [\jcap] {10.1088/1475-7516/2015/11/018}, \href
  {https://ui.adsabs.harvard.edu/abs/2015JCAP...11..018M} {2015, 018}

\bibitem[\protect\citeauthoryear{Michaux, Hahn, Rampf  \& Angulo}{Michaux
  et~al.}{2020}]{Michaux:2020yis}
Michaux M.,  Hahn O.,  Rampf C.,   Angulo R.~E.,  2020, \mn@doi [Mon. Not. Roy.
  Astron. Soc.] {10.1093/mnras/staa3149}, 500, 663

\bibitem[\protect\citeauthoryear{{Nadathur} \& {Crittenden}}{{Nadathur} \&
  {Crittenden}}{2016}]{Nadathur2016}
{Nadathur} S.,  {Crittenden} R.,  2016, \mn@doi [\apjl]
  {10.3847/2041-8205/830/1/L19}, \href
  {https://ui.adsabs.harvard.edu/abs/2016ApJ...830L..19N} {830, L19}

\bibitem[\protect\citeauthoryear{Nadathur \& Hotchkiss}{Nadathur \&
  Hotchkiss}{2015}]{Nadathur:2015qua}
Nadathur S.,  Hotchkiss S.,  2015, \mn@doi [Mon. Not. Roy. Astron. Soc.]
  {10.1093/mnras/stv2131}, 454, 2228

\bibitem[\protect\citeauthoryear{Nadathur, Lavinto, Hotchkiss  \&
  R\"as\"anen}{Nadathur et~al.}{2014a}]{PhysRevD.90.103510}
Nadathur S.,  Lavinto M.,  Hotchkiss S.,   R\"as\"anen S.,  2014a, \mn@doi
  [Phys. Rev. D] {10.1103/PhysRevD.90.103510}, 90, 103510

\bibitem[\protect\citeauthoryear{Nadathur, Hotchkiss, Diego, Iliev,
  Gottl\"ober, Watson  \& Yepes}{Nadathur et~al.}{2014b}]{Nadathur2016b}
Nadathur S.,  Hotchkiss S.,  Diego J.~M.,  Iliev I.~T.,  Gottl\"ober S.,
  Watson W.~A.,   Yepes G.,  2014b, \mn@doi [IAU Symp.]
  {10.1017/S1743921316010541}, 308, 542

\bibitem[\protect\citeauthoryear{Nadathur et~al.}{Nadathur
  et~al.}{2020}]{Nadathur:2020vld}
Nadathur S.,  et~al., 2020, \mn@doi [Mon. Not. Roy. Astron. Soc.]
  {10.1093/mnras/staa3074}, 499, 4140

\bibitem[\protect\citeauthoryear{{Neyrinck}}{{Neyrinck}}{2008}]{ZOBOV}
{Neyrinck} M.~C.,  2008, \mn@doi [\mnras] {10.1111/j.1365-2966.2008.13180.x},
  \href {https://ui.adsabs.harvard.edu/abs/2008MNRAS.386.2101N} {386, 2101}

\bibitem[\protect\citeauthoryear{Notari \& Riotto}{Notari \&
  Riotto}{2002}]{Notari:2002yc}
Notari A.,  Riotto A.,  2002, \mn@doi [Nucl. Phys. B]
  {10.1016/S0550-3213(02)00765-4}, 644, 371

\bibitem[\protect\citeauthoryear{O'Leary \& McQuinn}{O'Leary \&
  McQuinn}{2012}]{OLeary:2012}
O'Leary R.~M.,  McQuinn M.,  2012, \mn@doi [Astrophys. J.]
  {10.1088/0004-637X/760/1/4}, 760, 4

\bibitem[\protect\citeauthoryear{Obuljen, Villaescusa-Navarro, Castorina  \&
  Viel}{Obuljen et~al.}{2017}]{Obuljen:2016urm}
Obuljen A.,  Villaescusa-Navarro F.,  Castorina E.,   Viel M.,  2017, \mn@doi
  [JCAP] {10.1088/1475-7516/2017/09/012}, 09, 012

\bibitem[\protect\citeauthoryear{Odrzywo\l{}ek}{Odrzywo\l{}ek}{2009}]{PhysRevD.80.103515}
Odrzywo\l{}ek A.,  2009, \mn@doi [Phys. Rev. D] {10.1103/PhysRevD.80.103515},
  80, 103515

\bibitem[\protect\citeauthoryear{Peebles \& Hauser}{Peebles \&
  Hauser}{1974}]{peebles_1974}
Peebles P. J.~E.,  Hauser M.~G.,  1974, \mn@doi [Astrophysical Journal]
  {10.1086/190308}, 28, 19

\bibitem[\protect\citeauthoryear{Peloso, Pietroni, Viel  \&
  Villaescusa-Navarro}{Peloso et~al.}{2015}]{Peloso:2015jua}
Peloso M.,  Pietroni M.,  Viel M.,   Villaescusa-Navarro F.,  2015, \mn@doi
  [JCAP] {10.1088/1475-7516/2015/07/001}, 07, 001

\bibitem[\protect\citeauthoryear{Pezzotta, Crocce, Eggemeier, S\'anchez  \&
  Scoccimarro}{Pezzotta et~al.}{2021}]{Pezzotta:2021vfn}
Pezzotta A.,  Crocce M.,  Eggemeier A.,  S\'anchez A.~G.,   Scoccimarro R.,
  2021, \mn@doi [Phys. Rev. D] {10.1103/PhysRevD.104.043531}, 104, 043531

\bibitem[\protect\citeauthoryear{Philcox, Ivanov, Simonovi\'c  \&
  Zaldarriaga}{Philcox et~al.}{2020}]{Philcox:2020vvt}
Philcox O. H.~E.,  Ivanov M.~M.,  Simonovi\'c M.,   Zaldarriaga M.,  2020,
  \mn@doi [JCAP] {10.1088/1475-7516/2020/05/032}, 05, 032

\bibitem[\protect\citeauthoryear{Pisani, Sutter, Hamaus, Alizadeh, Biswas,
  Wandelt  \& Hirata}{Pisani et~al.}{2015a}]{PhysRevD.92.083531}
Pisani A.,  Sutter P.~M.,  Hamaus N.,  Alizadeh E.,  Biswas R.,  Wandelt B.~D.,
    Hirata C.~M.,  2015a, \mn@doi [Phys. Rev. D] {10.1103/PhysRevD.92.083531},
  92, 083531

\bibitem[\protect\citeauthoryear{{Pisani}, {Sutter}, {Hamaus}, {Alizadeh},
  {Biswas}, {Wandelt}  \& {Hirata}}{{Pisani} et~al.}{2015b}]{pisani2015}
{Pisani} A.,  {Sutter} P.~M.,  {Hamaus} N.,  {Alizadeh} E.,  {Biswas} R.,
  {Wandelt} B.~D.,   {Hirata} C.~M.,  2015b, \mn@doi [\prd]
  {10.1103/PhysRevD.92.083531}, \href
  {https://ui.adsabs.harvard.edu/abs/2015PhRvD..92h3531P} {92, 083531}

\bibitem[\protect\citeauthoryear{Platen, van~de Weygaert  \& Jones}{Platen
  et~al.}{2007}]{Platen:2007qk}
Platen E.,  van~de Weygaert R.,   Jones B. J.~T.,  2007, \mn@doi [Mon. Not.
  Roy. Astron. Soc.] {10.1111/j.1365-2966.2007.12125.x}, 380, 551

\bibitem[\protect\citeauthoryear{Platen, van~de Weygaert  \& Jones}{Platen
  et~al.}{2008}]{Platen:2007ng}
Platen E.,  van~de Weygaert R.,   Jones B. J.~T.,  2008, \mn@doi [Mon. Not.
  Roy. Astron. Soc.] {10.1111/j.1365-2966.2008.13019.x}, 387, 128

\bibitem[\protect\citeauthoryear{Polarski \& Starobinsky}{Polarski \&
  Starobinsky}{1994}]{Polarski:1994rz}
Polarski D.,  Starobinsky A.~A.,  1994, \mn@doi [Phys. Rev. D]
  {10.1103/PhysRevD.50.6123}, 50, 6123

\bibitem[\protect\citeauthoryear{Pollina, Baldi, Marulli  \&
  Moscardini}{Pollina et~al.}{2015}]{10.1093/mnras/stv2503}
Pollina G.,  Baldi M.,  Marulli F.,   Moscardini L.,  2015, \mn@doi [Monthly
  Notices of the Royal Astronomical Society] {10.1093/mnras/stv2503}, 455, 3075

\bibitem[\protect\citeauthoryear{Pollina, Hamaus, Dolag, Weller, Baldi  \&
  Moscardini}{Pollina et~al.}{2017}]{Pollina:2016gsi}
Pollina G.,  Hamaus N.,  Dolag K.,  Weller J.,  Baldi M.,   Moscardini L.,
  2017, \mn@doi [Mon. Not. Roy. Astron. Soc.] {10.1093/mnras/stx785}, 469, 787

\bibitem[\protect\citeauthoryear{{Raghunathan}, {Nadathur}, {Sherwin}  \&
  {Whitehorn}}{{Raghunathan} et~al.}{2020}]{Raghunathan2020}
{Raghunathan} S.,  {Nadathur} S.,  {Sherwin} B.~D.,   {Whitehorn} N.,  2020,
  \mn@doi [\apj] {10.3847/1538-4357/ab6f05}, \href
  {https://ui.adsabs.harvard.edu/abs/2020ApJ...890..168R} {890, 168}

\bibitem[\protect\citeauthoryear{Rampf, Uhlemann  \& Hahn}{Rampf
  et~al.}{2021}]{Rampf:2020ety}
Rampf C.,  Uhlemann C.,   Hahn O.,  2021, \mn@doi [Mon. Not. Roy. Astron. Soc.]
  {10.1093/mnras/staa3605}, 503, 406

\bibitem[\protect\citeauthoryear{Rees, Sciama  \& Stobbs}{Rees
  et~al.}{1968}]{Rees_1968}
Rees M.~J.,  Sciama D.~W.,   Stobbs S.~H.,  1968, Astrophysical Letters, 2, 243

\bibitem[\protect\citeauthoryear{Ricciardelli, Quilis  \&
  Planelles}{Ricciardelli et~al.}{2013}]{10.1093/mnras/stt1069}
Ricciardelli E.,  Quilis V.,   Planelles S.,  2013, \mn@doi [Monthly Notices of
  the Royal Astronomical Society] {10.1093/mnras/stt1069}, 434, 1192

\bibitem[\protect\citeauthoryear{Ricciardelli, Quilis  \& Varela}{Ricciardelli
  et~al.}{2014}]{10.1093/mnras/stu307}
Ricciardelli E.,  Quilis V.,   Varela J.,  2014, \mn@doi [Monthly Notices of
  the Royal Astronomical Society] {10.1093/mnras/stu307}, 440, 601

\bibitem[\protect\citeauthoryear{Schaap}{Schaap}{2007}]{Schapp-2007}
Schaap W.~E.,  2007, PhD Thesis - University of Groningen

\bibitem[\protect\citeauthoryear{Schmidt}{Schmidt}{2016}]{Schmidt:2016}
Schmidt F.,  2016, \mn@doi [Phys. Rev. D] {10.1103/PhysRevD.94.063508}, 94,
  063508

\bibitem[\protect\citeauthoryear{Schmidt}{Schmidt}{2021}]{Schmidt:2020tao}
Schmidt F.,  2021, \mn@doi [JCAP] {10.1088/1475-7516/2021/04/032}, 04, 032

\bibitem[\protect\citeauthoryear{{Schuster}, {Hamaus}, {Pisani}, {Carbone},
  {Kreisch}, {Pollina}  \& {Weller}}{{Schuster} et~al.}{2019}]{Schuster2019}
{Schuster} N.,  {Hamaus} N.,  {Pisani} A.,  {Carbone} C.,  {Kreisch} C.~D.,
  {Pollina} G.,   {Weller} J.,  2019, \mn@doi [\jcap]
  {10.1088/1475-7516/2019/12/055}, \href
  {https://ui.adsabs.harvard.edu/abs/2019JCAP...12..055S} {2019, 055}

\bibitem[\protect\citeauthoryear{{Sheth} \& {van de Weygaert}}{{Sheth} \& {van
  de Weygaert}}{2004}]{Sheth2004}
{Sheth} R.~K.,  {van de Weygaert} R.,  2004, \mn@doi [\mnras]
  {10.1111/j.1365-2966.2004.07661.x}, \href
  {https://ui.adsabs.harvard.edu/abs/2004MNRAS.350..517S} {350, 517}

\bibitem[\protect\citeauthoryear{Slepian \& Eisenstein}{Slepian \&
  Eisenstein}{2015}]{Slepian:2014dda}
Slepian Z.,  Eisenstein D.,  2015, \mn@doi [Mon. Not. Roy. Astron. Soc.]
  {10.1093/mnras/stu2627}, 448, 9

\bibitem[\protect\citeauthoryear{Slepian et~al.}{Slepian
  et~al.}{2018}]{Slepian:2016nfb}
Slepian Z.,  et~al., 2018, \mn@doi [Mon. Not. Roy. Astron. Soc.]
  {10.1093/mnras/stx2723}, 474, 2109

\bibitem[\protect\citeauthoryear{Springel}{Springel}{2005}]{Springel:2005mi}
Springel V.,  2005, \mn@doi [Mon. Not. Roy. Astron. Soc.]
  {10.1111/j.1365-2966.2005.09655.x}, 364, 1105

\bibitem[\protect\citeauthoryear{Sutter, Lavaux, Wandelt  \& Weinberg}{Sutter
  et~al.}{2012}]{Sutter_2012}
Sutter P.~M.,  Lavaux G.,  Wandelt B.~D.,   Weinberg D.~H.,  2012, \mn@doi [The
  Astrophysical Journal] {10.1088/0004-637x/761/2/187}, 761, 187

\bibitem[\protect\citeauthoryear{{Sutter}, {Lavaux}, {Hamaus}, {Wandelt},
  {Weinberg}  \& {Warren}}{{Sutter} et~al.}{2014a}]{Sutter2014}
{Sutter} P.~M.,  {Lavaux} G.,  {Hamaus} N.,  {Wandelt} B.~D.,  {Weinberg}
  D.~H.,   {Warren} M.~S.,  2014a, \mn@doi [\mnras] {10.1093/mnras/stu893},
  \href {https://ui.adsabs.harvard.edu/abs/2014MNRAS.442..462S} {442, 462}

\bibitem[\protect\citeauthoryear{Sutter, Pisani, Wandelt  \& Weinberg}{Sutter
  et~al.}{2014b}]{10.1093/mnras/stu1392}
Sutter P.~M.,  Pisani A.,  Wandelt B.~D.,   Weinberg D.~H.,  2014b, \mn@doi
  [Monthly Notices of the Royal Astronomical Society] {10.1093/mnras/stu1392},
  443, 2983

\bibitem[\protect\citeauthoryear{Taruya, Nishimichi, Saito  \&
  Hiramatsu}{Taruya et~al.}{2009}]{Taruya_2009}
Taruya A.,  Nishimichi T.,  Saito S.,   Hiramatsu T.,  2009, \mn@doi [Physical
  Review D] {10.1103/PhysRevD.80.123503}, 80

\bibitem[\protect\citeauthoryear{Tinker, Robertson, Kravtsov, Klypin, Warren,
  Yepes  \& Gottlober}{Tinker et~al.}{2010}]{Tinker:2010my}
Tinker J.~L.,  Robertson B.~E.,  Kravtsov A.~V.,  Klypin A.,  Warren M.~S.,
  Yepes G.,   Gottlober S.,  2010, \mn@doi [Astrophys. J.]
  {10.1088/0004-637X/724/2/878}, 724, 878

\bibitem[\protect\citeauthoryear{Tseliakhovich \& Hirata}{Tseliakhovich \&
  Hirata}{2010}]{Tseliakhovich_2010}
Tseliakhovich D.,  Hirata C.,  2010, \mn@doi [Physical Review D]
  {10.1103/physrevd.82.083520}, 82

\bibitem[\protect\citeauthoryear{Valiviita, Savelainen, Talvitie, Kurki-Suonio
  \& Rusak}{Valiviita et~al.}{2012}]{Valiviita:2012ub}
Valiviita J.,  Savelainen M.,  Talvitie M.,  Kurki-Suonio H.,   Rusak S.,
  2012, \mn@doi [Astrophys. J.] {10.1088/0004-637X/753/2/151}, 753, 151

\bibitem[\protect\citeauthoryear{{Verza}, {Pisani}, {Carbone}, {Hamaus}  \&
  {Guzzo}}{{Verza} et~al.}{2019}]{Verza2019}
{Verza} G.,  {Pisani} A.,  {Carbone} C.,  {Hamaus} N.,   {Guzzo} L.,  2019,
  \mn@doi [\jcap] {10.1088/1475-7516/2019/12/040}, \href
  {https://ui.adsabs.harvard.edu/abs/2019JCAP...12..040V} {2019, 040}

\bibitem[\protect\citeauthoryear{Viel, Colberg  \& Kim}{Viel
  et~al.}{2008}]{Viel_2008}
Viel M.,  Colberg J.~M.,   Kim T.-S.,  2008, \mn@doi [Monthly Notices of the
  Royal Astronomical Society] {10.1111/j.1365-2966.2008.13130.x}, 386, 1285

\bibitem[\protect\citeauthoryear{{Vielzeuf} et~al.,}{{Vielzeuf}
  et~al.}{2021}]{vielzeuf2021}
{Vielzeuf} P.,  et~al., 2021, \mn@doi [\mnras] {10.1093/mnras/staa3231}, \href
  {https://ui.adsabs.harvard.edu/abs/2021MNRAS.500..464V} {500, 464}

\bibitem[\protect\citeauthoryear{Villaescusa-Navarro, Alonso  \&
  Viel}{Villaescusa-Navarro et~al.}{2017}]{Villaescusa-Navarro:2016kbz}
Villaescusa-Navarro F.,  Alonso D.,   Viel M.,  2017, \mn@doi [Mon. Not. Roy.
  Astron. Soc.] {10.1093/mnras/stw3224}, 466, 2736

\bibitem[\protect\citeauthoryear{Yoo \& Seljak}{Yoo \&
  Seljak}{2013}]{Yoo:2013qla}
Yoo J.,  Seljak U.,  2013, \mn@doi [Phys. Rev. D] {10.1103/PhysRevD.88.103520},
  88, 103520

\bibitem[\protect\citeauthoryear{Yoo, Dalal  \& Seljak}{Yoo
  et~al.}{2011}]{Yoo_2011}
Yoo J.,  Dalal N.,   Seljak U.,  2011, \mn@doi [Journal of Cosmology and
  Astroparticle Physics] {10.1088/1475-7516/2011/07/018}, 2011, 018–018

\bibitem[\protect\citeauthoryear{Yoshida, Sugiyama  \& Hernquist}{Yoshida
  et~al.}{2003}]{Yoshida:2003}
Yoshida N.,  Sugiyama N.,   Hernquist L.,  2003, \mn@doi [Mon. Not. Roy.
  Astron. Soc.] {10.1046/j.1365-8711.2003.06829.x}, 344, 481

\bibitem[\protect\citeauthoryear{Zeldovich}{Zeldovich}{1970}]{Zeldovich-1970}
Zeldovich Y.~B.,  1970, \mn@doi [A\&A] {1970A&A.....5...84Z}, 500, 13

\bibitem[\protect\citeauthoryear{Zivick, Sutter, Wandelt, Li  \& Lam}{Zivick
  et~al.}{2015}]{10.1093/mnras/stv1209}
Zivick P.,  Sutter P.~M.,  Wandelt B.~D.,  Li B.,   Lam T.~Y.,  2015, \mn@doi
  [Monthly Notices of the Royal Astronomical Society] {10.1093/mnras/stv1209},
  451, 4215

\makeatother
\end{thebibliography}




\bsp	
\label{lastpage}
\end{document}